\documentclass[aps,superscriptaddress,twocolumn]{revtex4-1}

\pdfoutput=1

\usepackage{graphicx}
\usepackage{amsmath}
\usepackage{amssymb}
\usepackage{verbatim}
\usepackage{braket}
\usepackage{color}
\usepackage{hyperref}
\hypersetup{
	colorlinks=true,
	linkcolor=red,      
	urlcolor=blue,
	citecolor=blue
}

\begin{document}
\title{Topological transition between gapless phases in quantum walks}
	
\author{Ranjith R Kumar}
\affiliation{Department of Physics, Indian Institute of Technology Bombay, Powai, Mumbai 400076, India.}
\author{Hideaki Obuse}
\affiliation{Department of Applied Physics, Hokkaido University, Sapporo 060-8628, Japan.}

\date{\today}

\begin{abstract}
Topological gapless phases of matter have been a recent interest among theoretical and experimental condensed matter physicists. Fermionic chains with extended nearest neighbor couplings have been observed to show unique topological transition at the multicritical points between distinct gapless phases. In this work, we show that such topological gapless phases and the transition between them can be simulated in a quantum walk. We consider a three-step discrete-time quantum walk and identify various critical or gapless phases and multicriticalities from the topological phase diagram along with their distinguished energy dispersions. We reconstruct the scaling theory based on the curvature function to study transition between gapless phases in the quantum walk. We show the interesting features observed in fermionic chains, such as diverging, sign flipping and swapping properties of curvature function, can be simulated in the quantum walk. Moreover, the renormalization group flow and Wannier state correlation functions also identify transition at the multicritical points between gapless phases. We observe the scaling law and overlapping of critical and fixed point properties at the multicritical points of the fermionic chains can also be observed in the quantum walk. Furthermore, we categorize the topological transitions at various multicritical points using the group velocity of the energy eigenstates. Finally, the topological characters of various gapless phases are captured using winding number which allows one to distinguish various gapless phases and also show the transitions at the multicritical points.

\end{abstract}

\maketitle

\section{Introduction}
Topological aspects of quantum matter have been receiving the limelight in condensed matter physics for the past two decades \cite{haldane1988model,kane2005quantum,hasan2010colloquium,narang2021topology,wang2017topological}. The topological properties embedded in the electronic band structure produce novel phases of quantum matter which can be characterized by topological invariant numbers \cite{thouless1982quantized}. The distinct gapped phases are associated with quantized topological numbers that change at the transition point which involves bulk band gap closing. Therefore, no topological properties are associated at the critical/gapless point between distinct gapped phases\cite{continentino2020finite}.

However, recently, this conventional notion has been carefully re-examined in a few extended topological chains. Interestingly, analytical and numerical investigations show that the non-trivial topological properties even at gapless phases \cite{verresen2018topology,verresen2019gapless,jones2019asymptotic,verresen2020topology,rahul2021majorana,niu2021emergent,PhysRevB.104.075132,PhysRevResearch.3.043048,fraxanet2021topological,keselman2015gapless,scaffidi2017gapless,duque2021topological,kumar2021multi,kwwangSciPostPhys.12.4.134,kumar2021topological,kumar2023topological}.
As the conventional topological invariant number is ill-defined at the critical points, the non-triviality of the gapless phases can be characterized in terms of the zero and poles of complex function associated with the Hamiltonian \cite{verresen2018topology}.
The symmetry properties of the low-energy conformal field theory also provide topological invariants for gapless systems \cite{verresen2019gapless}. Moreover, topological properties of gapless systems were characterized using the correlations of string operators \cite{jones2019asymptotic}, conformal boundary conditions \cite{PhysRevLett.129.210601}, universal entanglement spectrum \cite{yu2024universal} etc.
The generalization of the gapless phases in multiband models with chiral symmetry, extended quantum XXZ models, and gapless Floquet systems have also been studied \cite{PhysRevResearch.3.043048,PhysRevA.109.062226,cardoso2024gapless,zhou2025gapless}. The topological characters of these gapless phases are robust against the disorders and interactions \cite{duque2021topological,keselman2015gapless}. 

In topological models with extended nearest-neighbor couplings, a unique toplogical transition between gapless phases have been characterized based on a \textit{near-critical approach} where the system is critical only in parameter space while away from its critical momentum \cite{kumar2021topological}. The transition between these distinct gapless phases occurs through a multicritical point without bulk gap opening. Such gapless phases were distinguished into high symmetry and non-high symmetry based on the gap closing points in the Brillouin zone and are characterized using the winding number, curvature function, entanglement entropy, etc \cite{kumar2021topological,kumar2023topological}.

On the other hand, various topological quantum matter has been simulated in varieties of experimental platforms such as superconducting circuits \cite{PhysRevB.101.035109,niu2021simulation}, ultracold atoms in optical lattices \cite{goldman2016topological,meier2016observation} etc. Recently, quantum walks were found to be effective simulators for various topological systems due to the underlying symmetries \cite{PhysRevA.103.012201,PhysRevA.82.033429}. The edge states of topological systems of all the symmetry classes in 1D and 2D have been simulated with good controllability in quantum walks. It also allows one to obtain the topological invariants 
and identify the topological phase transition between distinct gapped phases \cite{PhysRevB.88.121406,PhysRevLett.118.130501}. In particular, topological gapped phases of SSH models in 1D and 2D have been simulated using discrete-time quantum walks. This enabled one to realize the symmetries, edge modes, and topological invariants in these systems. Moreover, a scaling theory based on the curvature function have been implimented to analyse the topological transition between the gapped phases in 1D and 2D quantum walks \cite{panahiyan2020fidelity}.

However, so far topological gapped phases have been simulated in quantum walks  
whilst topological gapless phases and transition between them remains unexplored. 
In this work, we explore the unique topological transition at the multicritical points between gapless phases in a quantum walk. We also develop the scaling theory, based on the curvature function, to understand such transitions and identify the topological characters of the gapless phases.

The paper is laid out as follows. In Section.\ref{sec2}, we introduce the model and discuss the details of its energy dispersion, eigenvalue distribution, etc. In Section.\ref{sec3}, we calculate the winding number for various topological gapped phases and obtain the topological phase diagram. We identify the corresponding phase boundaries or criticalities and multicriticalities. 
Gapless topological phases of the quantum walk and the transition between them are discussed in Section.\ref{sec4}. We construct the scaling theory based on the curvature function which involves the behavior of curavture function, critical exponents, renormalization group flow, Wannier state correlation function and group velocity in characterizing the unique topological transition between gapless phases.
In Section.\ref{sec5}, we identify the topological characters of various gapless phases of the quantum walk in terms of winding numbers. We also discuss the nature of winding of unit vectors at the gapless phases. Finally, we summarize and conclude in Section.\ref{sec6}.

\section{Model}\label{sec2}
Topological transition between distinct gapless phases have been observed in topological chains with two or more nearest-neighbor couplings \cite{kumar2021topological,kumar2023topological}. As these models can generate higher winding number phases and multicritical points which separate distinct gapless phases, we look for appropriate quantum walks to simulate such systems \cite{hideakimodel,higherwqw}. Therefore, we consider discrete-time three-step unitary quantum walk over a one-dimensional lattice. Since the number of steps in quantum walks mimics the nearest neighbor couplings on a lattice, the three-step quantum walk can generate high winging number phases.

\begin{figure}[t]
	\includegraphics[width=4.2cm,height=2.8cm]{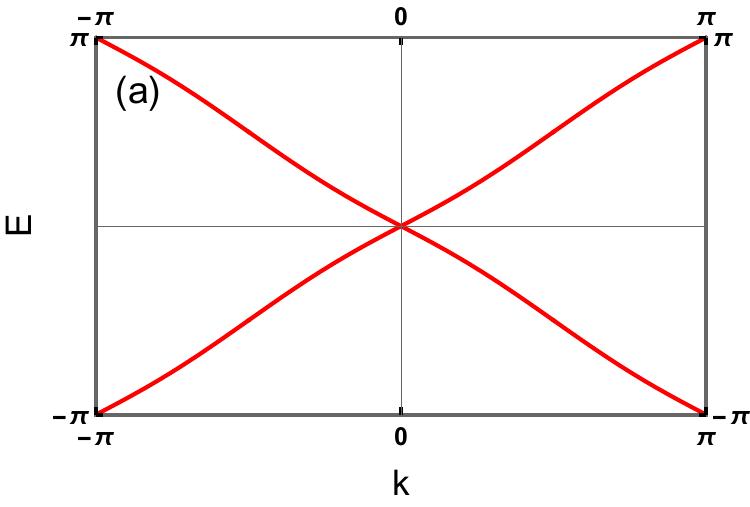}\hspace{0.2cm}\includegraphics[width=4.2cm,height=2.8cm]{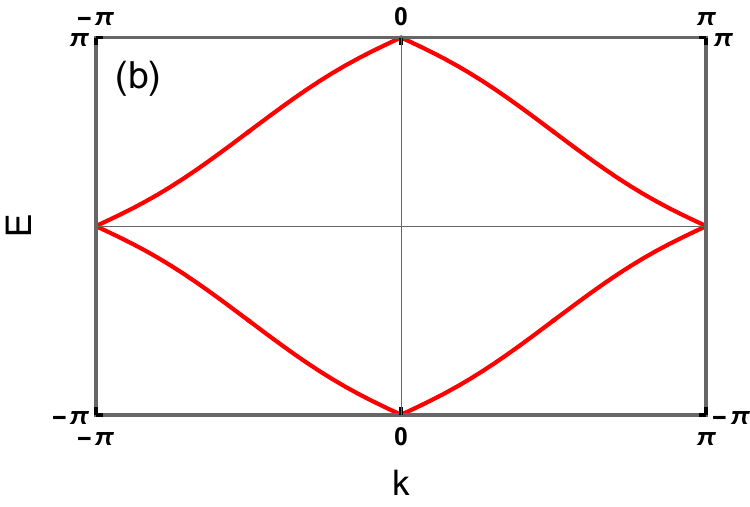}\\
	\includegraphics[width=4.2cm,height=2.8cm]{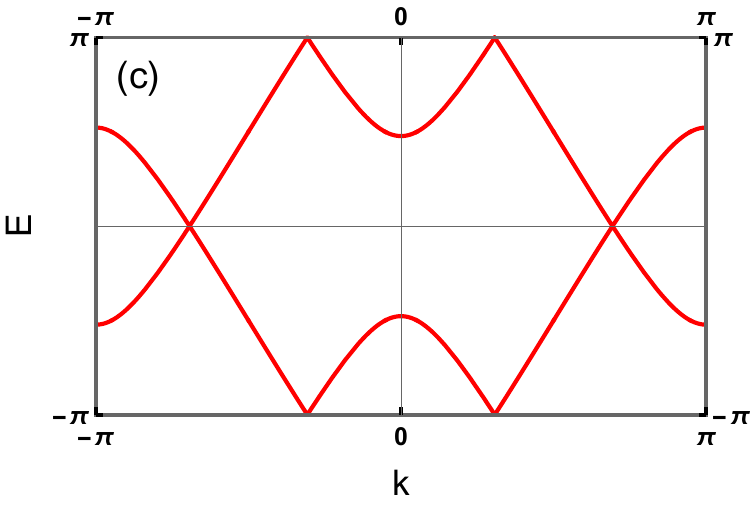}\hspace{0.2cm}\includegraphics[width=4.2cm,height=2.8cm]{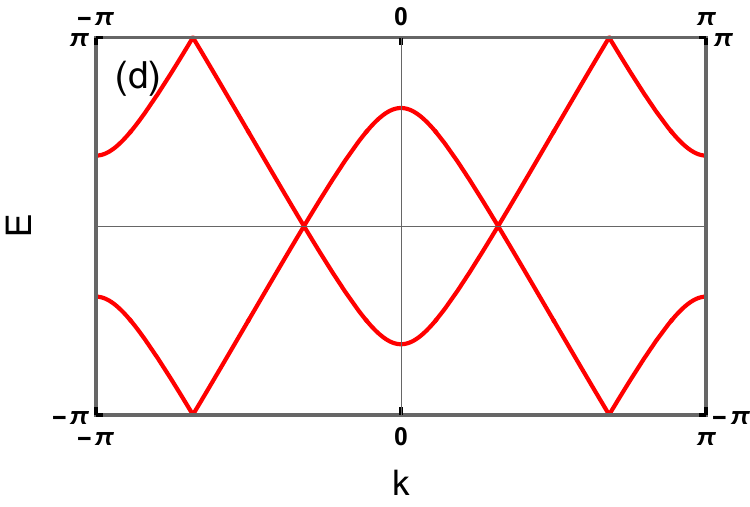}\\
	\caption{Dispersion of three-step quantum walk in Eq.\ref{disp}. Gap closing occurs at $E=0$ and $E=\pm \pi$ for (a) $k_0=0$ and $k_0=\pm \pi$ respectively, (b) $k_0=\pm \pi$ and $k_0=0$ respectively. In (c) and (d) the gap closing occurs at arbitrary values of $k$.}
	\label{1disp}
\end{figure}

The time-evolution operator of such a system can be defined as 
\begin{equation}
U_3= SC[\theta_2(x)]SC[\theta_2(x)]SC[\theta_1(x)],
\end{equation}
where $S$ is shift operator and $C[\theta_j(x)]$ ($j=1,2$) are coin operators, such that
\begin{equation}
S=\sum_{x} \left( \left|x-1\right\rangle \left\langle x\right| \otimes \left|L\right\rangle \left\langle L\right| + \left|x+1\right\rangle \left\langle x\right| \otimes \left|R\right\rangle \left\langle R\right|\right) \nonumber
\end{equation}
\begin{equation}
C[\theta_j(x)] = \sum_{x} \left| x\right\rangle \left\langle x\right| \otimes \begin{pmatrix}
\cos \theta_j(x) & -\sin \theta_j(x)\\
\sin \theta_j(x) & \cos \theta_j (x)
\end{pmatrix} \nonumber
\end{equation}
where $x\in \mathbb{Z}$ is the position of the walker with the internal degrees of freedom $R$ and $L$. The coin operator $C[\theta_j(x)]$ is the rotation operator in the coin space and acts on the internal states of the systems. The shift operator $S$ moves the walker's position to the right or left side based on its internal states.

To realize the symmetries of the model, one can redefine $U_3$ in a symmetric time frame. 
\begin{equation}
U^{\prime}_3= C[\frac{\theta_1(x)}{2}]SC[\theta_2(x)]SC[\theta_2(x)]SC[\frac{\theta_1(x)}{2}].
\end{equation}
The diagonalization of $U^{\prime}_3$ in the Fourier space yields,
\begin{equation}
U^{\prime}_3(k) = d_0(k)I+d_1(k)\sigma_1+id_2(k)\sigma_2+id_3(k)\sigma_3 \label{3sqw}
\end{equation}
with 
\begin{align}
d_0(k)&= -(\cos(\theta_1)\sin^2(\theta_2)+\sin(\theta_1)\sin(2\theta_2))\cos(k) \nonumber\\&+ \cos(\theta_1)\cos^2(\theta_2) \cos(3k) \nonumber\\
d_1(k)&= 0 \nonumber\\
d_2(k)&= (\sin(\theta_1)\sin^2(\theta_2)-\cos(\theta_1)\sin(2\theta_2))\cos(k) \nonumber\\&- \sin(\theta_1)\cos^2(\theta_2) \cos(3k) \nonumber\\
d_3(k)&= -\sin^2(\theta_2)\sin(k)+\cos^2(\theta_2) \sin(3k)
\end{align}
Stroboscopically, the evolution operator can be mapped to an effective Hamiltonian as 
\begin{equation}
H(k)= i \ln U^{\prime}_3 = E \left[ \mathbf{n(k).\sigma}\right]
\end{equation}
where $E$ is the quasi-energy dispersion and $\mathbf{\sigma}$ are Pauli spin matrices. The winding vector $\mathbf{n(k)}$ can be described as 
\begin{equation}
\mathbf{n(k)}= \frac{d_1}{\sqrt{d_1^2+d_2^2+d_3^2}},\frac{d_2}{\sqrt{d_1^2+d_2^2+d_3^2}},\frac{d_3}{\sqrt{d_1^2+d_2^2+d_3^2}}.
\end{equation} 
The energy dispersion can be defined as $E=i \ln \lambda$, where $\lambda= d_0(k)\pm i \sqrt{1-d_0(k)^2}$ is the eigenvalue of the evolution operator. 
For Eq.\ref{3sqw}, the eigenvalue and dispersion can be obtained as
\begin{align}
\lambda &= \alpha \pm \frac{1}{4} \sqrt{-2\beta_1 - 8 \sin^2(2k) \beta_2 +2 \cos^2(k) \beta_3} \label{eigen}\\
E&=\pm \cos^{-1}(\alpha)\label{disp}\\
\text{where,} \nonumber
\end{align}

\begin{align}
\alpha&=\cos(3 k) \cos(\theta_1) \cos(\theta_2)^2 \nonumber \\&- 
\cos(k) \sin(\theta_2) (2 \cos(\theta_2) \sin(\theta_1) + 
\cos(\theta_1) \sin(\theta_2)), \nonumber\\
\beta_1&=(9+4\cos(2k)+3\cos(4k))\sin^2(k)\cos(2\theta_1), \nonumber	\\
\beta_2&=2\cos(2k)\cos^2(\theta_1)\cos(2\theta_2)+\sin(2\theta_1)\sin(2\theta_2), \nonumber\\
\beta_3&=9+3\cos(4k)+(2\cos(4k)\cos^2(\theta_1)-5\cos(2\theta_1)-1)\nonumber\\
&\cos(4\theta_2)+4\cos(2k)(\sin(2\theta_1)\sin(4\theta_2)-2)	
\end{align}
The effective Hamiltonian has chiral (sub-lattice) symmetry defined as $\Gamma^{-1}H(k)\Gamma=-H(k)$, where $\Gamma=\sigma_1$ is the chiral symmetry operator. As a consequence of this symmetry, we obtain a symmetric spectrum, i.e. $E$ and $-E$ always appear in pairs. Due to the periodicity of quasi-energy, the gap closing ($E = -E$) occurs at $E = 0$ and $E = \pm\pi$, as shown in Fig.\ref{1disp}. 
In this model, the gap closing occurs for the high symmetry points in the Brillouin zone i.e. $k_0=0$ and $k_0=\pm \pi$ as well as non-high symmetry points (arbitrary values) of $k$. This distinction enables us to identify the multicritical points (discussed later) which act as the topological transition points between gapless phases.

\begin{figure}
	\begin{center}
		\includegraphics[width=7cm,height=7cm]{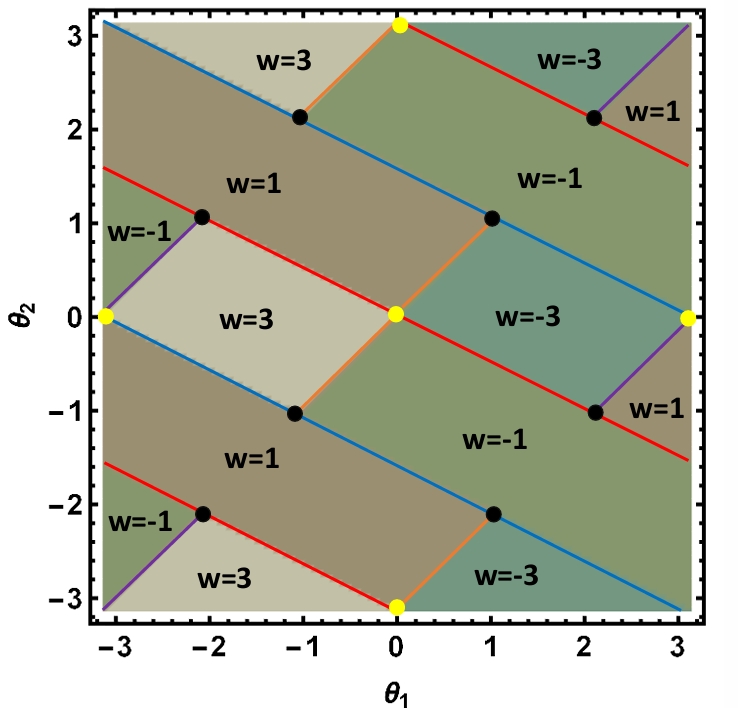}
	\end{center}
	\caption{Topological phase diagram of three-step quantum walk. Various gapped phases are pinned with the respective values of $w$. The phase boundaries or criticalities are marked in different colors based on the gap-closing points in the Brillouin zone. The multicritical points are marked black and yellow which shows quadratic and linear dispersions respectively.}
	\label{phase}
\end{figure}

\section{Topological phase diagram}\label{sec3}
\begin{figure}
	\includegraphics[width=6cm,height=4cm]{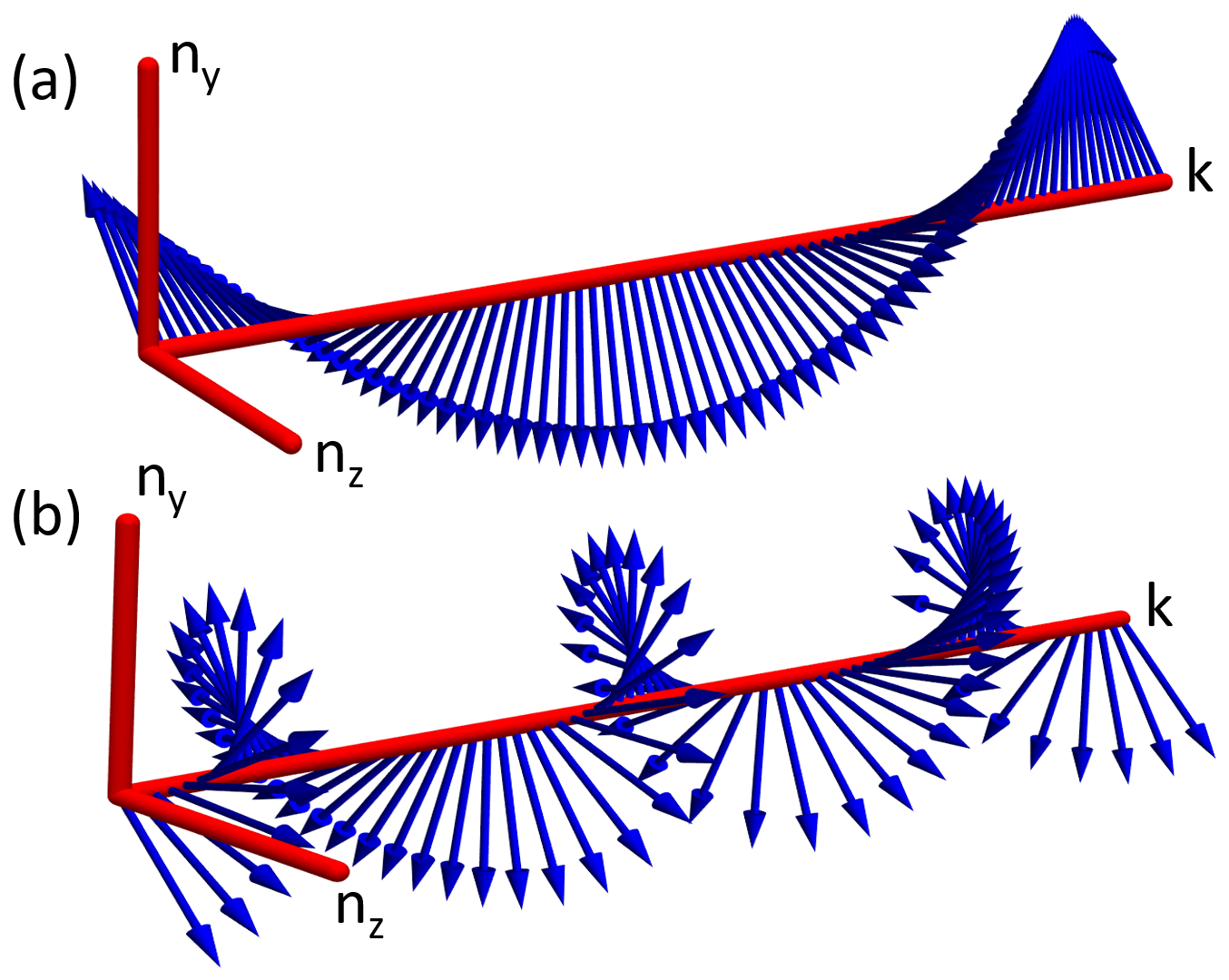} 
	\caption{Winding vector $\mathbf{n(k)}$ of three-step quantum walk. (a) Shows the the topological gapped phases with $w=\pm 1$ and (b) shows the topological gapped phases with $w=\pm 3$.}
	\label{WWG}
\end{figure}

The topological properties of the three-step quantum walk 
can be identified by calculating the winding number. It can be expressed as \cite{hideakimodel} 
\begin{equation}
w=\frac{1}{2\pi} \int_{-\pi}^{\pi} \frac{d\theta_k}{dk} dk, \label{winding}
\end{equation}
where $\theta_k=\tan^{-1}(d_3/d_2)$. 
The topological phase diagram can be obtained based on the values of $w$, as shown in Fig. \ref{phase}, which can identify different topological gapped phases and the topological phase transitions between them.
The winding number $w$ represents the winding of unit vector $\mathbf{n(k)}$ around the origin. In other words, $\mathbf{n(k)}$ traces a closed path around the origin as $k$ goes from $-\pi$ to $\pi$ \cite{PhysRevLett.115.177204}. The model hosts distinct gapped phases with $w=\pm 1$ and $w=\pm 3$. For a gapped phase with $w=\pm 1$, the winding of the unit vector has single winding in the Brillouin zone, as shown in Fig. \ref{WWG}(a). Similarly, for $w=\pm 3$ one can identify three windings around the origin, as shown in Fig. \ref{WWG}(b).

These phases are separated by the line of critical points (critical line) where the gap closes at $E=0$ and $E=\pm \pi$. The critical lines (colored red, blue, orange, and purple) are distinguished based on the nature of dispersion and gap-closing points. 
The critical lines in the topological phase diagrams can be obtained as,
\begin{equation}
\text{\underline{Red lines}:} \nonumber
\end{equation}
\begin{gather}
(\theta_1,\theta_2)=(0,\pi) \leftrightarrow (\theta_1,\theta_2)=(\pi,\pi/2)  \nonumber\\
\theta_2=-\left( \frac{\theta_1-2\pi}{2}\right) \label{R1}
\end{gather}
\begin{gather}
(\theta_1,\theta_2)=(-\pi,\pi/2) \leftrightarrow (\theta_1,\theta_2)=(\pi,-\pi/2)  \nonumber\\
\theta_2=-\left( \frac{\theta_1}{2}\right) \label{R2}
\end{gather}
\begin{gather}
(\theta_1,\theta_2)=(-\pi,-\pi/2) \leftrightarrow (\theta_1,\theta_2)=(0,-\pi)  \nonumber\\
\theta_2=-\left( \frac{\theta_1+2\pi}{2}\right) \label{R3}
\end{gather}

These are the high symmetry critical lines where gap closing occurs at $E=0$ for $k_0=0$ and $E=\pm \pi$ for $k_0=\pm \pi$. 
\begin{equation}
\text{\underline{Blue lines}:} \nonumber
\end{equation}
\begin{gather}
(\theta_1,\theta_2)=(-\pi,\pi) \leftrightarrow (\theta_1,\theta_2)=(\pi,0)  \nonumber\\
\theta_2=-\left( \frac{\theta_1-\pi}{2}\right) \label{B1}
\end{gather}
\begin{gather}
(\theta_1,\theta_2)=(-\pi,0) \leftrightarrow (\theta_1,\theta_2)=(\pi,-\pi)  \nonumber\\
\theta_2=-\left( \frac{\theta_1+\pi}{2}\right) \label{B2}
\end{gather}
These are the high symmetry critical lines where gap closing occurs at $E=0$ for $k_0=\pm \pi$ and $E=\pm \pi$ for $k_0=0$.
\begin{equation}
\text{\underline{Orange and Purple lines}:} \nonumber
\end{equation}
\begin{gather}
(\theta_1,\theta_2)=(-\pi,-\pi) \leftrightarrow (\theta_1,\theta_2)=(\pi,\pi)  \nonumber\\
\theta_2=\theta_1 \label{OP1}
\end{gather}
\begin{gather}
(\theta_1,\theta_2)=(-\pi,0) \leftrightarrow (\theta_1,\theta_2)=(0,\pi)  \nonumber\\
\theta_2=(\theta_1+\pi) \label{OP2}
\end{gather}
\begin{gather}
(\theta_1,\theta_2)=(0,-\pi) \leftrightarrow (\theta_1,\theta_2)=(\pi,0)  \nonumber\\
\theta_2=(\theta_1-\pi) \label{OP3}
\end{gather}
These are the non-high symmetry critical lines where gap closing occurs at $E=0$ and $E=\pm \pi$ for arbitrary values of $k$ except for the high symmetry points.

The red critical line shows the dispersion shown in Fig. \ref{1disp}(a). Similarly, blue, orange, and purple critical lines refer to the kind of dispersion in Fig. \ref{1disp}(b), (c), and (d) respectively. Therefore, red and blue lines can be referred to as high symmetry critical lines whilst orange and purple lines are non-high symmetry lines \cite{murakami2011gap,kourtis2017weyl,molignini2018universal,kumar2023topological}. These critical lines set the boundaries of various gapped phases and facilitates the topological phase transition between distinct phases. This transition can be identified with a gap closing and a change in winding number.

\begin{figure}[t]
	\includegraphics[width=4cm,height=3cm]{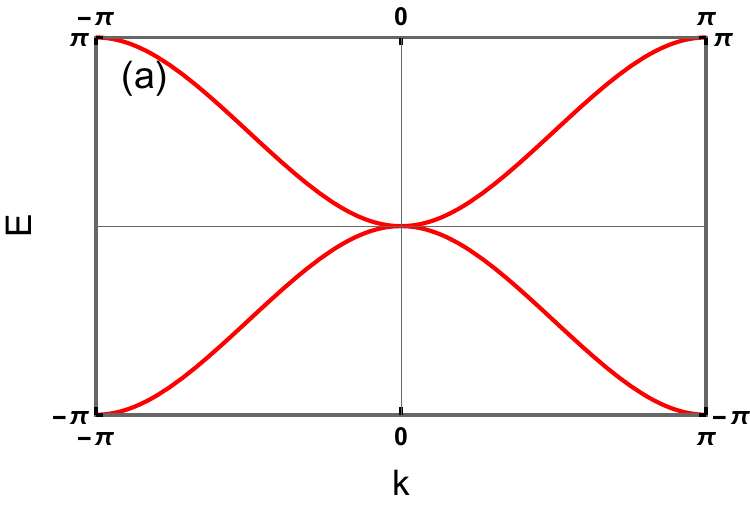}\hspace{0.5cm}\includegraphics[width=4cm,height=3cm]{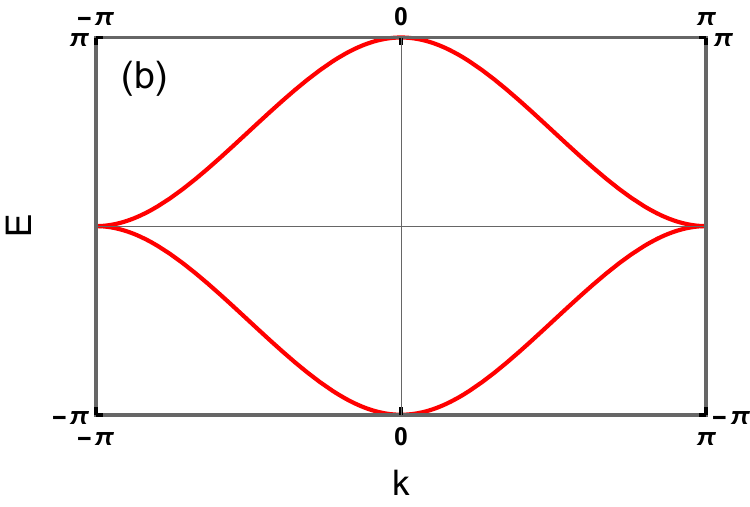}\\
	\includegraphics[width=4cm,height=3cm]{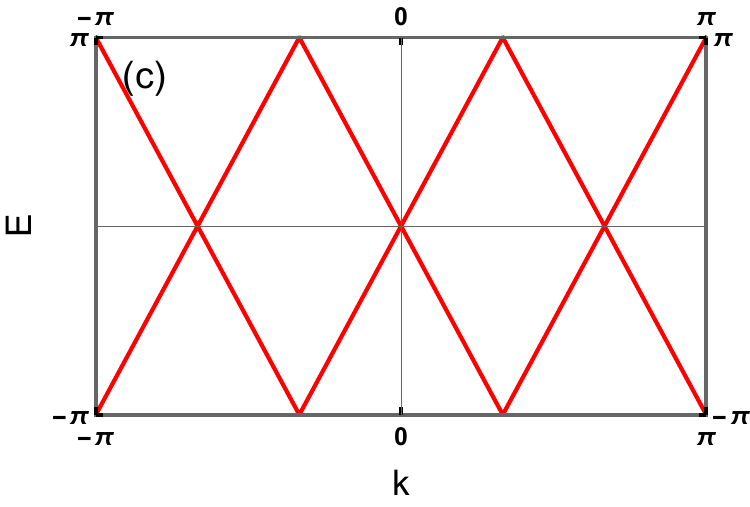}\hspace{0.5cm}\includegraphics[width=4cm,height=3cm]{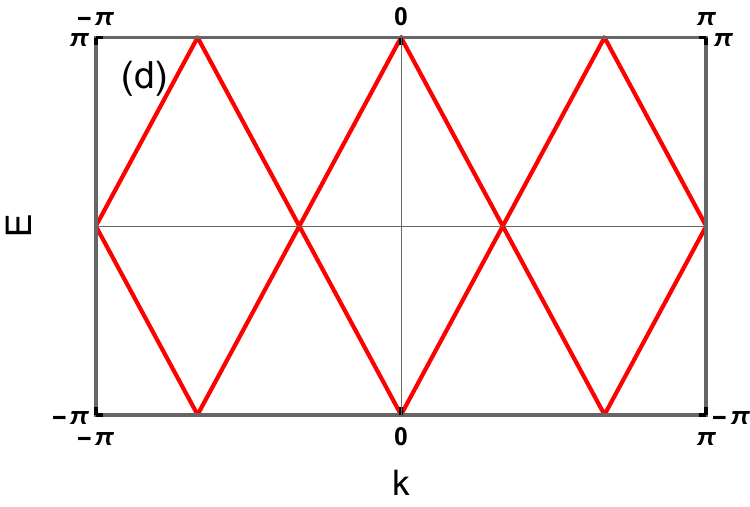}
	\caption{Energy dispersion at the multicritical points. (a) Quadratic dispersion at the points in Eq.\ref{MC1} with $\theta_1=\pm\frac{2\pi}{3}$. (b) Quadratic dispersion at the points in Eq.\ref{MC1} with $\theta_1=\pm\frac{\pi}{3}$. (c) Linear dispersion at the points in Eq.\ref{MC2} with $\theta_1=0$, (d) Linear dispersion at the points in Eq.\ref{MC2} with $\theta_1=\pm\pi$.}
	\label{DQL}
\end{figure}

Moreover, there are several special points, referred to as multicritical points, which are the intersection points of two critical lines. These points differ from the other critical points due to their nature of dispersion, as shown in Fig.\ref{DQL}. Based on this they can be categorized into two kinds. The first kind (black dots) possess quadratic dispersion, as shown in Fig.\ref{DQL}(a) and (b), and the second kind (yellow dots) possess linear dispersion, as shown in Fig.\ref{DQL}(c) and (d). 
\begin{equation}
\text{\underline{Black dots}:} \nonumber
\end{equation}
\begin{gather}
\left( \theta_1,\theta_2\right) =\left( -\frac{2\pi}{3},-\frac{2\pi}{3}\right), \left( -\frac{2\pi}{3},\frac{\pi}{3}\right), \left( -\frac{\pi}{3},-\frac{\pi}{3}\right),\nonumber\\
 \left( -\frac{\pi}{3},\frac{2\pi}{3}\right), 
\left( \frac{\pi}{3},-\frac{2\pi}{3}\right), \left( \frac{\pi}{3},\frac{\pi}{3}\right),\nonumber\\
\left( \frac{2\pi}{3},-\frac{\pi}{3}\right), \left( \frac{2\pi}{3},\frac{2\pi}{3}\right).
\label{MC1}
\end{gather}

\begin{equation}
\text{\underline{Yellow dots}:} \nonumber
\end{equation}
\begin{gather}
\left(\theta_1,\theta_2\right)=\left(0,0\right) , \left(\pi,0\right), \left(-\pi,0\right),
\left(0,\pi\right), 
\left(0,-\pi\right).
\label{MC2}
\end{gather}
These points divide a critical line into several segments which can be referred to as gapless or critical phases and, therefore, mediates a transition between these gapless phases along the line. This unique phenomenon is discussed in the next section. 

\section{Topological transition between gapless phases}\label{sec4}
Apart from the various topological gapped phases, one can also identify topologically distinct gapless phases in the three-step quantum walk considered in this work. The multicritical points marks the topological transition between such gapless phases. However, these transitions does not involve bulk gap closing and opening as in the conventional case discussed previously. Therefore, this phenomenon can not be captured using conventionally methods as most physical quantities diverge at a critical point. Nevertheless, in the topological chains, a near-critical approach has been proposed that captures this unique phenomenon without referring to any of the gapped phases of the model \cite{kumar2021topological,kumar2023topological}. We adopt the same approach here to address the topological transition between gapless phases of the quantum walk.

Conventionally, driving the system to criticality requires both parameter and momentum at its critical values. In the near critical approach, we keep quantum walk critical only in parameter space with $k = k_0 + \Delta k$, where $k_0$ is the critical value of momentum and $\Delta k\ll2\pi$ \cite{kumar2021topological}. This approach avoids the singularities of an exact gapless point and enables one to redefine various characterizing tools to capture the transition between gapless phases at the multicritical points. We implement this by considering the relations between the parameters $\theta_1$ and $\theta_2$, of the quantum walk for all the critical lines i.e. Eq.\ref{R1}-Eq.\ref{OP3}.

\begin{figure}[t]
	\includegraphics[width=4.2cm,height=2.8cm]{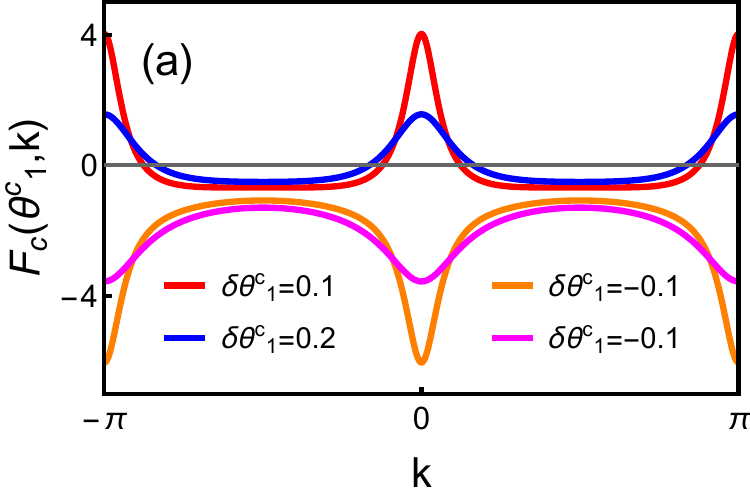}\hspace{0.2cm}\includegraphics[width=4.2cm,height=2.8cm]{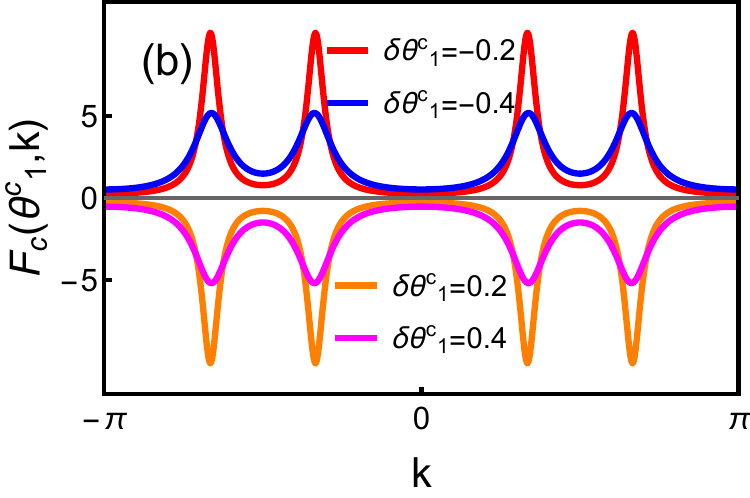}\\
	\includegraphics[width=4.2cm,height=2.8cm]{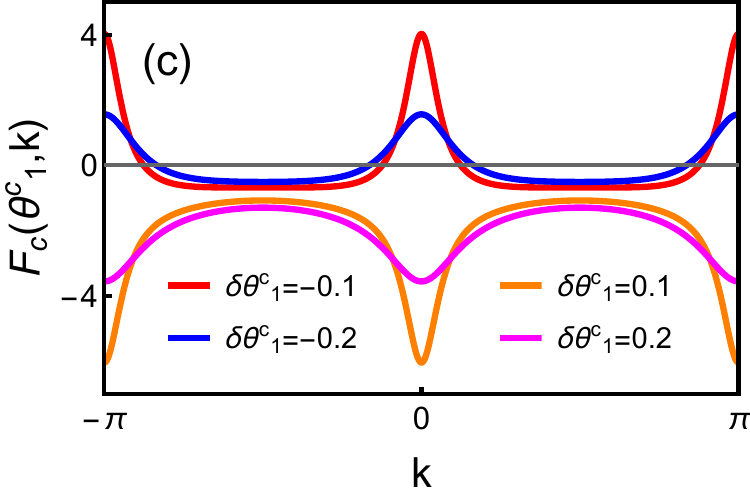}\hspace{0.2cm}\includegraphics[width=4.2cm,height=2.8cm]{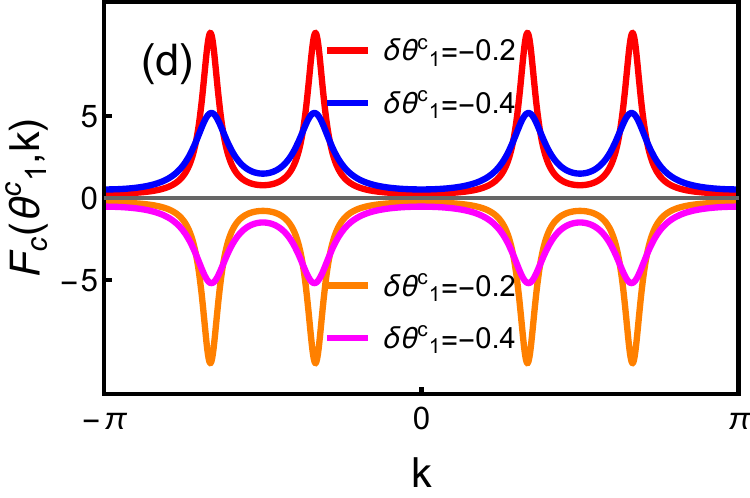}\\
	\includegraphics[width=4.2cm,height=2.8cm]{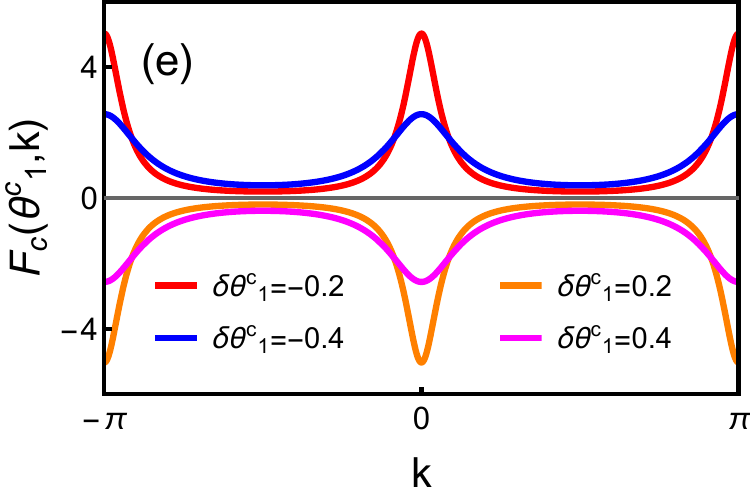}\hspace{0.2cm}\includegraphics[width=4.2cm,height=2.8cm]{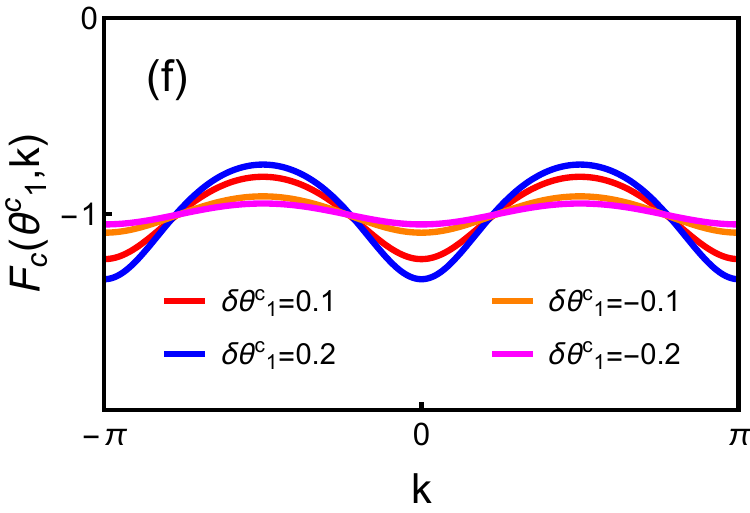}
	\caption{Profile of the curvature function in the vicinity of multicritical points. (a) Near $\theta_1^{mc}=\pm\frac{2\pi}{3}$ and (b) near $\theta_1^{mc}=0$ on the red critical lines in Eq.\ref{R1}-Eq.\ref{R3}. (c) Near $\theta_1^{mc}=\pm\frac{\pi}{3}$ and (d) near $\theta_1^{mc}=\pm\pi$ on the blue critical lines in Eq.\ref{B1}-Eq.\ref{B2}. (e) Near $\theta_1^{mc}=0$ and (f) near $\theta_1^{mc}=\pm\frac{2\pi}{3},\pm\frac{\pi}{3}$ on the orange and purple critical lines in Eq.\ref{OP1}-Eq.\ref{OP3}. The diverging peak and sign flip of the curvature function indicate the topological transition between gapless phases at these multicritical points.}
	\label{CF}
\end{figure}
\subsection{Curvature function and critical exponents}
In general, topological and critical properties of gapped phases are encapsulated in the curvature function \cite{rufo2019multicritical,kartik2021topological,chen2016scaling}.  Topology of a given phase can be quantified upon integration of the curvature function over the first Brillouin zone, which gives the quantized invariant numbers (such as winding number in Eq.\ref{winding}). Whilst, the critical behavior can be identified with the diverging nature of curvature function in the vicinity of a topological phase transition point. A set critical exponents associated with this divergence defines the universality class of the critical point \cite{chen2017correlation}. These signatures of curvature function can also be generalized for gapless phases using the near-critical approach \cite{kumar2021topological,kumar2023topological}. The transition between gapless phases exhibit similar diverging nature of curvature function at the multicritical points. In this subsection we discuss such behaviors in the case of three-step quantum walk considered in this work.

The curvature function of the three-step quantum walk at criticality can be obtained by substituting for $\theta_2$ according to the criticality conditions defined in Eq.\ref{R1}-Eq.\ref{OP3}. Therefore, the gapless quantum walk is tuned using only one parameter $\theta_1^{c}$ which goes along a critical line through mulicritical points. The curvature function  
can be obatined as
\begin{equation}
F(\theta_1^{c},k) = \frac{d^c_2\partial_k d^c_3-d^c_3\partial_k d^c_2}{(d^c_2)^2+(d^c_3)^2}
\end{equation}
where $d^c_3$ and $d^c_2$ are the components of the Hamiltonian on a critical line. 
For example, to address the topological transition between gapless phases on a red critical line in Eq.\ref{R1}, we substitute Eq.\ref{R1} in $d_3$ and $d_2$ which yields $d^c_3$ and $d^c_2$ respectively. 

\begin{figure}
	\includegraphics[width=4.2cm,height=2.8cm]{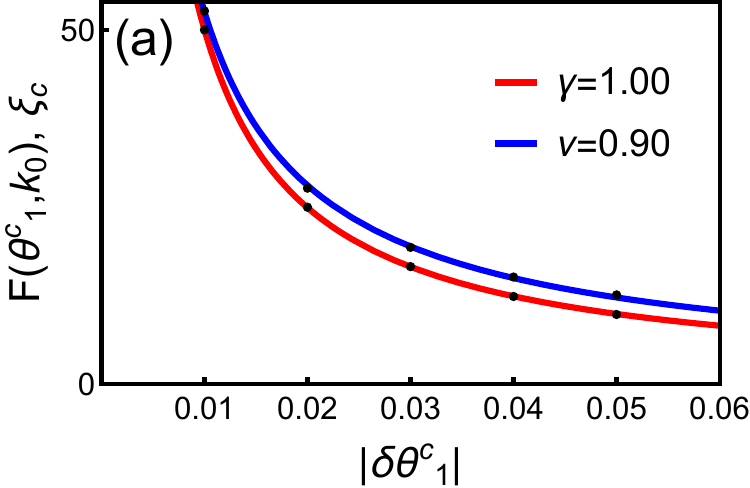}\hspace{0.2cm}\includegraphics[width=4.2cm,height=2.8cm]{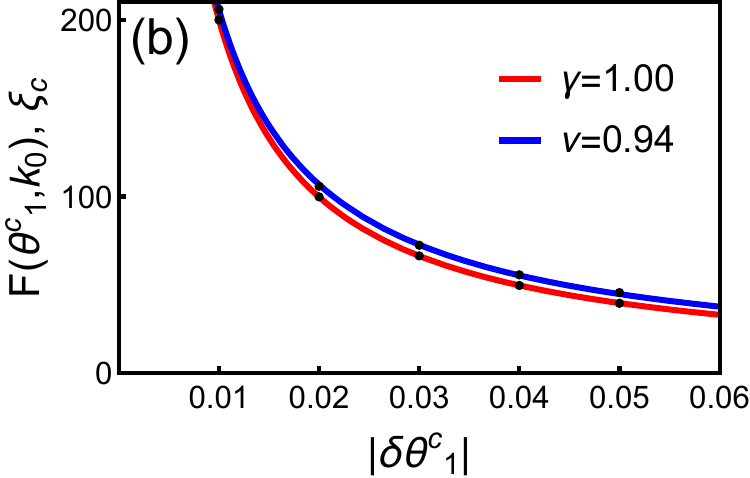}\\
	\includegraphics[width=4.2cm,height=2.8cm]{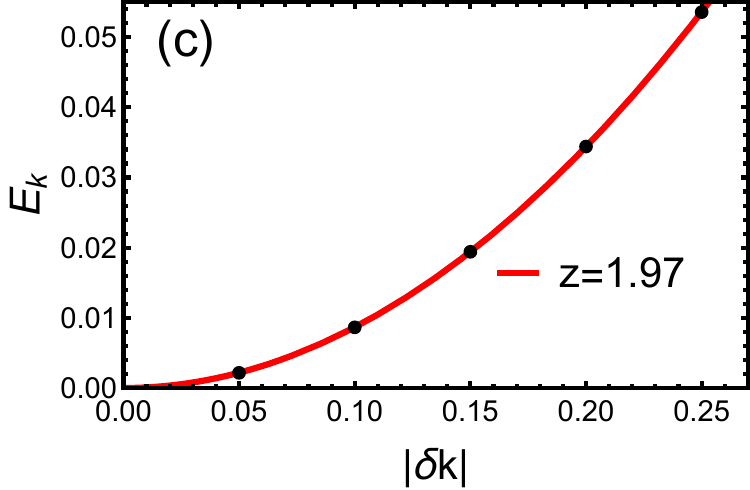}\hspace{0.2cm}\includegraphics[width=4.2cm,height=2.8cm]{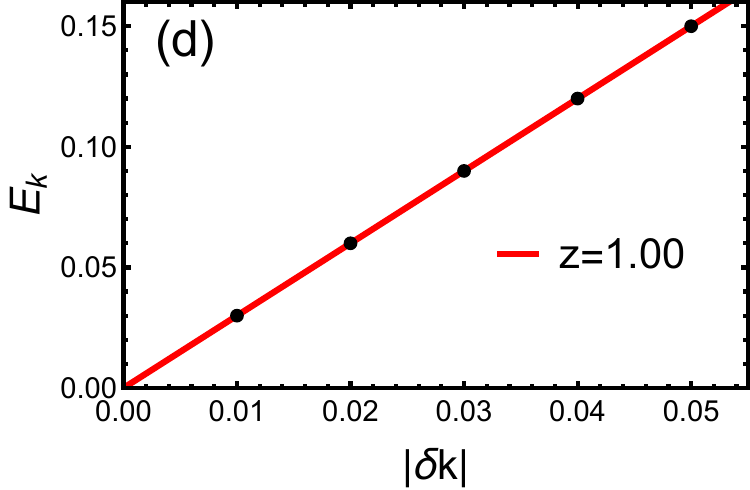}
	\caption{Critical exponents of curvature function and energy dispersion. (a) Exponents of the curvature function at the multicritical points $\theta_1^{mc}=0,\pm\frac{2\pi}{3},\pm\frac{\pi}{3}$ (i.e. shown in Fig.\ref{CF}(a),(c),(e)). (b) Exponents of the curvature function at the multicritical points $\theta_1^{mc}=0,\pm\pi$ (i.e. shown in Fig.\ref{CF}(b),(d)). In both cases we observe $\gamma=\nu=1$. (c) Dynamical exponent at the multicritical points $\theta_1^{mc}=\pm\frac{2\pi}{3}, \pm\frac{\pi}{3}$ shows $z=2$. (d) Dynamical exponent at the multicritical points $\theta_1^{mc}=0,\pm \pi$ shows $z=1$.}
	\label{criexp}
\end{figure}

As the parameter space approaches multicritical points along a critical line i.e. $\theta_1^c\rightarrow\theta_1^{mc}$, the curvature function develops a diverging peak at $k_0$ with $F(\theta_1^c,k_0+\delta k)=F(\theta_1^c,k_0-\delta k)$ where $\delta k$ is small  deviation away from the $k_0$. Moreover, the sign of the diverging peak flips across a multicritical points indicating a transition between gapless phases,
\begin{equation}
\lim_{\theta_1^{c}\rightarrow \theta_1^{mc+}} F(\theta_1^{c},k_0)=-\lim_{\theta_1^{c}\rightarrow \theta_1^{mc-}} F(\theta_1^{c},k_0)=\pm\infty
\end{equation}
\begin{figure*}[t]
	\includegraphics[width=5cm,height=3.5cm]{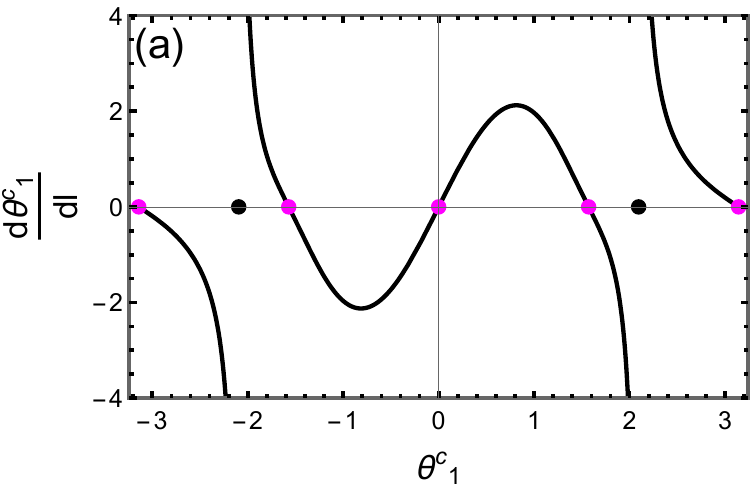}\hspace{0.4cm}\includegraphics[width=5cm,height=3.5cm]{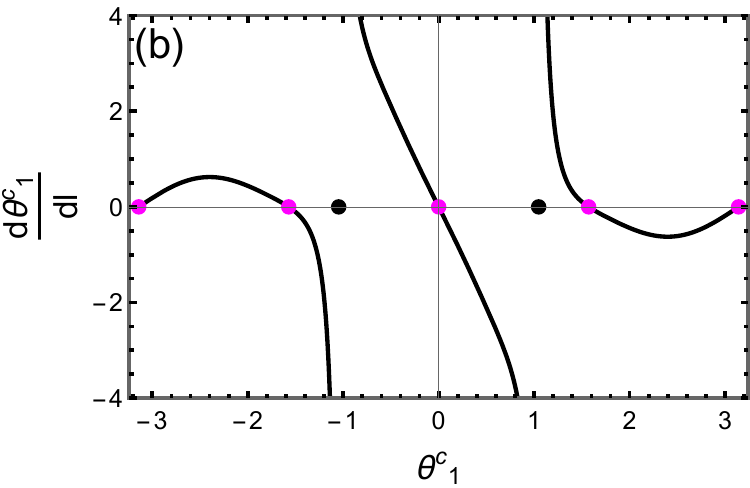}\hspace{0.4cm}\includegraphics[width=5cm,height=3.5cm]{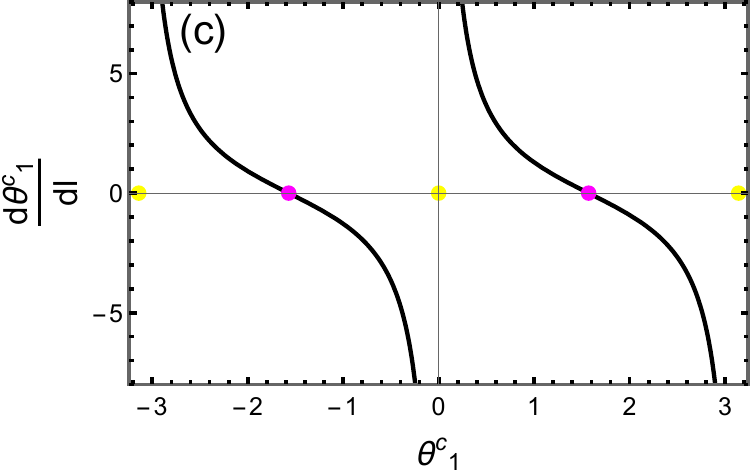}
	\caption{Curvature function RG flow at various criticalities. (a) For red critical lines in Eq.\ref{R1}-Eq.\ref{R3}, where the transition occurs at $\theta_1^{mc}=\pm\frac{2\pi}{3}$. (b) For blue critical lines in Eq.\ref{B1}-Eq.\ref{B2}, where the transition occurs at $\theta_1^{mc}=\pm\frac{\pi}{3}$. (c) For orange and purple critical lines in Eq.\ref{OP1}-Eq.\ref{OP3}, where the transition occurs at $\theta_1^{mc}=0,\pm \pi$. The points where the flow goes to zero $\theta_{1f}^{c}=0,\pm\frac{\pi}{2},\pm \pi$ , are the fixed points.}
	\label{RG}
\end{figure*}
The diverging curve obtain Ornstein-Zernike form around $k_0$,
\begin{equation}
F(\theta_1^c,k_0+\delta k)=\frac{F(\theta_1^c,k_0)}{1+\xi_c^2\delta k^2}
\label{OZ}
\end{equation}
where, $\xi_c$ is the characteristic length scale at
criticality that represents the width of the curvature function around a multicritical point. The critical properties can be quantified in terms of critical exponents $\gamma$ and $\nu$ as
\begin{equation}
F(\theta_1^{c},k_0)\propto |\theta_1^{c}-\theta_1^{mc}|^{-\gamma}, \;\;\;\;\;\; \xi_c \propto |\theta_1^{c}-\theta_1^{mc}|^{-\nu}
\label{ce}
\end{equation}
These critical exponents characterize the universality class of multicritical points.
 
At first, we consider critical lines (red lines) in Eq.\ref{R1}-Eq.\ref{R3} and identify the topological transition of the respective gapless phases. At the multicritical points $\theta_1^c=\theta_1^{mc}$, the gap closing occurs at $E=0$ and $E=\pm\pi$ at $k_0=0$ and $k_0=\pm\pi$ respectively. The curvature functions for these critical lines can be obtained as

\begin{equation}
F(\theta_1^{c},k)=\frac{-\cos^3\left( \frac{\theta_1^{c}}{2}\right) \sin\left( \frac{\theta_1^{c}}{2}\right)  \eta^{r}_{1}}{25+7\eta^{r}_{2}+16\eta^{r}_{3}+\cos(2k) \eta^{r}_{4}- \sin^2(2k) \eta^{r}_{5}}
\label{CFR}
\end{equation}
where, $\eta^{r}_{1}=128(1 + 2 \cos^2(k) \cos(\theta_1^{c}))$, $\eta^{r}_{2}=\cos (4k)$, $\eta^{r}_{3}=\cos^2(k)\cos(\theta_1^{c})$, $\eta^{r}_{4}=16[1 + \cos^2(k) (3 \cos\theta_1^{c} + 2 \cos(2\theta_1^{c}))]$ and $\eta^{r}_{5}=2[4 \cos(3 \theta_1^{c}) + \cos(4 \theta_1^{c})]$. There are three multicritical points on these critical lines at $\theta_1^{mc}=0,\pm 2\pi/3$ which are characterized by linear and quadratic dispersions respectively. We show that the curvature functions in Eq.\ref{CFR} diverge at these multicritical points indicating topological transition between gapless phases on the critical lines. In Fig.\ref{CF}, evolution of the curvature function in the Brillouin in the vicinity of the multicritical points is shown. The curvature function develops a diverging peak (i) at $k_0=0,\pm \pi$ for $\theta_1^{mc}=\pm 2\pi/3$, as shown in Fig.\ref{CF}(a) and (ii) at non-high symmetry $k_0$ points for $\theta_1^{mc}=0$, as shown in Fig.\ref{CF}(b). Moreover, the sign of curvature function flips across the multicritical points indicating the topological transitions. Note that, in Fig.\ref{CF}(b), missing diverging peaks at $k_0=0,\pm \pi$ are due to the swaping property of the curvature function at the multicritical points \cite{kumar2021topological}, which is discussed later.

Now we consider the critical lines (blue lines) in Eq.\ref{B1} and Eq.\ref{B2} where the gap closing occurs at $E=0$ and $E=\pm\pi$ at $k_0=\pm \pi$ and $k_0=0$ respectively. We obtain the curvature function for these critical lines as
\begin{equation}
F(\theta_1^{c},k)=\frac{\cos^3\left( \frac{\theta_1^{c}}{2}\right) \sin\left( \frac{\theta_1^{c}}{2}\right)  \eta^{b}_{1}}{25+7\eta^{b}_{2}-16\eta^{b}_{3}+\cos(2k) \eta^{b}_{4}-\sin^2(2k)\eta^{b}_{5}}
\end{equation}
where, $\eta^{b}_{1}=128(-1 + 2 \cos^2(k) \cos(\theta_1^{c}))$, $\eta^{b}_{2}=\cos(4k)$, $\eta^{b}_{3}=\cos^2(k)\cos(\theta_1^{c})$, $\eta^{b}_{4}=16[1 - \cos^2(k) (3 \cos\theta_1^{c} - 2 \cos(2\theta_1^{c}))]$ and $\eta^{b}_{5}=2[-4 \cos(3 \theta_1^{c}) - \cos(4 \theta_1^{c})]$. In this case, the multicirtical points with quadratic dispersion is obtained at $\theta_1^{mc}=\pm\pi/3$ and linear dispersive points are at $\theta_1^{mc}=\pm\pi$. As the parameter $\theta_1^c$ runs towards $\theta_1^{mc}=\pm\pi/3$ the diverging peak develops at high symmetry points $k_0=0,\pm \pi$, as shown in Fig.\ref{CF}(c). Similarly, near $\theta_1^{mc}=\pm\pi$ the diverging peak develops at the non-high symmetry $k_0$ points, as shown in Fig.\ref{CF}(d). As in the previous case, topological transitions can be identified from the sign change across the multicritical points. 

On the critical lines (purple and orange lines) in Eq.\ref{OP1}-Eq.\ref{OP3}, gap closes at $E=0,\pm\pi$ for non-high symmetry $k_0$ points. The curvature function in these cases can be obtained as
\begin{equation}
F(\theta_1^{c},k)=\frac{4 \sin(\theta_1^{c})}{\cos(2k)+2 \cos^2(k) \cos(2\theta_1^{c})-3}
\end{equation}
The multicriticalities are at $\theta_1^{mc}=0,\pm\pi/3,\pm2\pi/3$. Among them only $\theta_1^{mc}=0$ has linear dispersion while other four points are characterized by quadratic dispersions. Fig.\ref{CF}(e) shows the curvature function near $\theta_1^{mc}=0$ with the diverging peak with a sign flip. However, the curvature function shows no interesting feature near quadratic multicritical points at $\theta_1^{mc}=\pm\pi/3,\pm2\pi/3$, as shown in Fig.\ref{CF}(f). This is due to the fact that only $\theta_1^{mc}=0$ has gapless phases on both sides along the critical lines, and therefore it hosts topological transition between these gapless phases. The multicritical points $\theta_1^{mc}=\pm\pi/3,\pm2\pi/3$ dictates a phase boundary between gapless and gapped phases along the critical lines. Therefore, the curvature function remain insensitive to these points and indicate no such topological transitions between gapless phases can be observed. 

Note that, interestingly the swaping property of the curvature function during gapless-gapless transition in fermionic systems \cite{kumar2021topological} can also be simulated in quantum walks. The curvature function near the linear multicritical points has gap closing at both $k_0=0,\pm\pi$ and intermediate $k_0$ values since they are the intersection points of high and non-high symmetry critical lines. Therefore, the diverging peak is expected to appear at all $k_0$ points. However, along the high symmetry critical lines the peak appear at only non-high symmetry $k_0$ points, as shown in Fig.\ref{CF}(b) and (d). Similarly, along the non-high symmetry lines the peak appear at only high symmetry points $k_0=0,\pm\pi$, as shown in Fig.\ref{CF}(e). This swaping of $k_0$ points appear as a consequence of intersection of the critical lines. We do not observe such swaping of $k_0$ points at quadratic multicritical points, because the non-high symmetry $k_0$ points confluence with high symmetry $k_0$ points at these multicriticalities resulting in quadratic dispersions. 

The multicritical properties are quantified and the universality classes can be obtained from the critical exponents $\gamma$ and $\nu$, as in Eq.\ref{ce}. At first, we expand components of the Hamiltonian $d_2^c$ and $d_3^c$, on the high symmetry critical lines, around the gap closing $k_0=0,\pm \pi$ upto third order,
\begin{align}
d_2^c(k)|_{k=k_0}&\approx \zeta_1 \delta k^2 \nonumber\\
d_3^c(k)|_{k=k_0}&\approx \zeta_2 \delta k+\zeta_3 \delta k^3
\end{align}
where 
\begin{align}
\zeta_1&=\frac{1}{2}(\pm 9\cos^2\theta_2 \sin\theta_1 \mp \cos\theta_1 \sin\theta_1 \mp \sin^2\theta_2 \sin\theta_1) \nonumber\\
\zeta_2&=\pm 3 \cos^2\theta_2 \mp \sin^2\theta_2 \nonumber\\
\zeta_3&=\frac{1}{6}(\mp 27 \cos^2\theta_2 \pm \sin^2\theta_2) \nonumber\\
\end{align}
Here the signs are respectively for $k_0=0,\pm \pi$ and $\theta_2$ will be substituted by the corresponding critical lines in Eq.\ref{R1}-Eq.\ref{B2}. 
Now the curvature function can be written in the Ornstein-Zernike form in Eq.\ref{OZ} as
\begin{align}
F(\theta_1^c,k_0+\delta k)&=  \frac{d^c_2\partial_k d^c_3-d^c_3\partial_k d^c_2}{(d^c_2)^2+(d^c_3)^2}\nonumber\\
&= \frac{\left( \frac{\zeta_1 \zeta_2 \delta k^2 - \zeta_1 \zeta_3 \delta k^4}{\zeta_2^2 \delta k^2} \right) }{1+\left( \frac{\zeta_1^2+2 \zeta_2 \zeta_3}{\zeta_2^2} \right)\delta k^2 } \nonumber\\
&=\frac{F(\theta_1^c,k_0)}{1+\xi_c^2\delta k^2}.
\end{align}
The dominant behavior in $F(\theta_1^c,k_0)$ and $\xi_c$ leads to $F(\theta_1^c,k_0)=\xi_c=\zeta_1\zeta_2^{-1}$. Therefore, the critical exponents for both multicritical points are $\gamma=\nu=1$. These exponents can also be calculated by numerically fitting the 
curvature function around $k_0$ to the Ornstein-Zernike form in Eq.\ref{OZ}. In Fig.\ref{criexp}(a) and Fig.\ref{criexp}(b) we show the critical exponents obtained for the multicritical points $\theta_1^{mc}=\pm\frac{2\pi}{3},\pm\frac{\pi}{3}$ and $\theta_1^{mc}=0,\pm\pi$ respectively. In both cases we observe the exponents are estimated to be $\gamma=\nu\sim1$ which agrees with the analytical values. Therefore, the scaling law that $\gamma=\nu$ in one dimensional fermionic systems can be simulated in three-step quantum walk. Moreover, the quadratic and linear nature of dispersion at the multicritical points are also shown in Fig.\ref{criexp}(c) and Fig.\ref{criexp}(d) respectively. This yields the dynamical critical exponent $z$ (defined as $E\propto k^z$) to be $z=2$ (quadratic) and $z=1$ (linear) which differentiates the universality classes of the multicritical points.

\subsection{Renormalization group flow and correlation function}
The diverging curvature function in the vicinity of transition points allows one to construct a scaling theory \cite{chen2016scaling,chen2016scalinginvariant,chen2017correlation,chen2018weakly,chen2019universality,molignini2018universal,panahiyan2020fidelity,molignini2020generating,abdulla2020curvature,malard2020scaling,molignini2020unifying,kumar2021multi}.  The diverging peak can be gradually reduced by driving the parameter space towards a fixed point configuration of the curvature function. This leads to a renormalization group (RG) flow equation which identify the critical and fixed points in the parameter space. This theory has been extended to the quantum walks recently \cite{panahiyan2020fidelity}. The diverging curvature function, associated critical exponents, scaling laws and the stroboscopic Wannier state correlation function have been calculated for a simple quantum walk that simulate topological transitions. Here we extend this idea to the gapless phases of the three-step quantum walk.

For a given parameter space $\theta_1^{c}$, close to $\theta_1^{mc}$, a diverging peak of the curvature function can be reduced by finding a new $\theta_1^{c\prime}$ away from $\theta_1^{mc}$, such that 
\begin{equation}
F(\theta_1^{c\prime},k_0)= F(\theta_1^{c},k_0+\delta k)
\end{equation} 
where $\delta k$ is small  deviation away from the $k_0$. Therefore, the scaling drives the system to its fixed point configuration $F(\theta_{1f}^{c},k)$ without changing the topology.  Performing the scaling procedure iteratively one can obtain a RG equation 
with the scaling parameters $\delta k^2=dl$ and $|\theta_1^{c\prime}-\theta_1^{c}|=d\theta_1^{c}$.
\begin{equation}
\frac{d\theta_1^{c}}{dl}=\frac{1}{2}\frac{\partial^2_k F(\theta_1^{c},k)|_{k=k_0}}{\partial_{\theta_1^{c}} F(\theta_1^{c},k_0)}
\label{CRG}
\end{equation}
The topological transition between gapless phases can be identified with divergence in the RG flow at the multicritical points. 
Moreover, the fixed points can be identified with vanishing RG flow 
in the parameter space. 

\begin{figure}[t]
	\includegraphics[width=4.2cm,height=2.8cm]{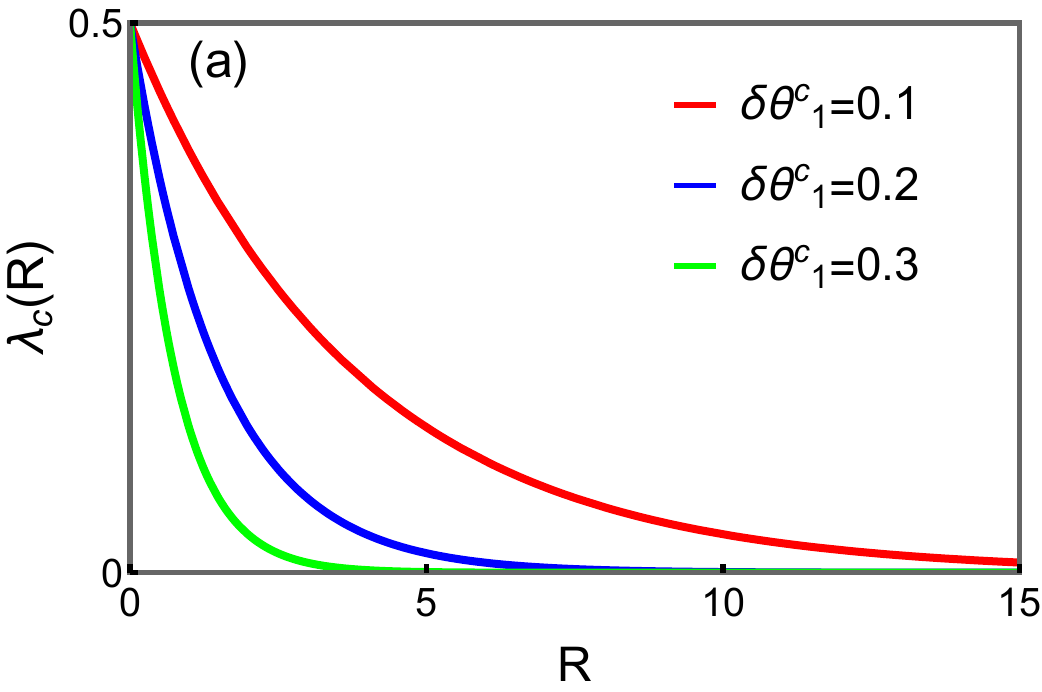}\hspace{0.2cm}\includegraphics[width=4.2cm,height=2.8cm]{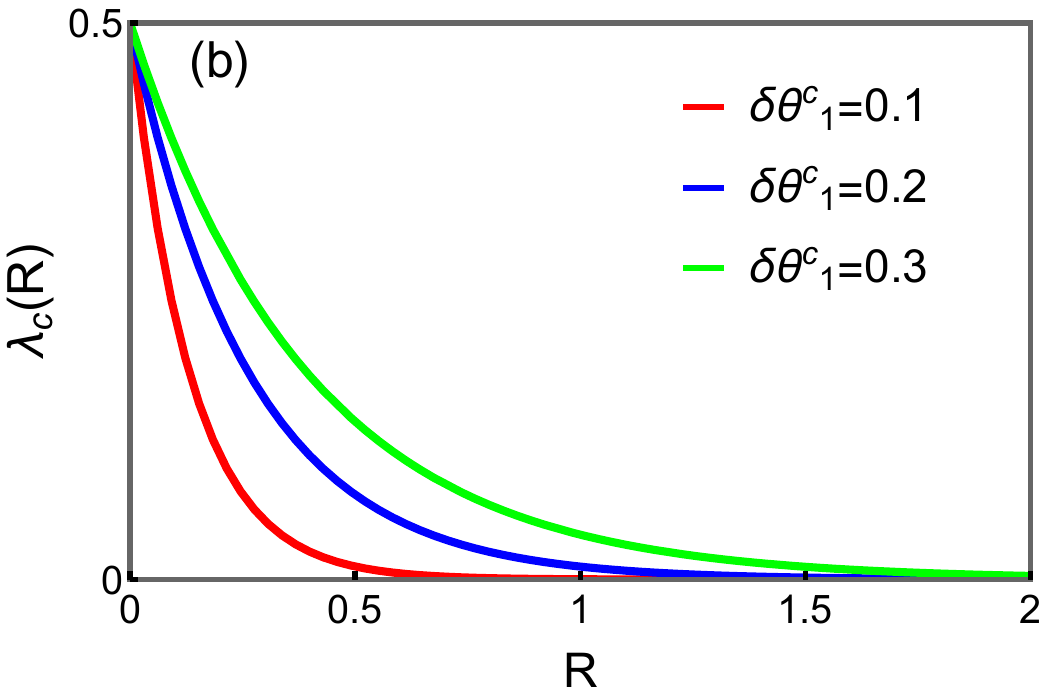}\
	\caption{Wannier state correlation function in the vicinity of multicritical points. As the distance from the $\theta_1^{mc}$ increased by $\delta\theta_1^{c}$ the corresponding decay of $\lambda_c(R)$ with $R$ is shown. (a) Near $\theta_1^{mc}=\pm\frac{2\pi}{3}, \pm\frac{2\pi}{3}$. With increasing $\delta\theta_1^{c}$ sharp decay of $\lambda_c(R)$ can be observed. (b) Near $\theta_1^{mc}=0,\pm\pi$. With increasing $\delta\theta_1^{c}$ slow decay of $\lambda_c(R)$ can be observed.}
	\label{WCF}
\end{figure}

Furthermore, a correlation function of the Wannier states at different positions can be obtained from the Fourier transform of the curvature function which identify the topological transition between gapless phases of quantum walk \cite{panahiyan2020fidelity}. Therefore, the correlation function of the Wannier states at a distance $R$ can be obtained as
\begin{equation}
\lambda_c(R)=e^{i k_0 R}\frac{F(\theta_1^{c},k_0)}{2\xi_c} e^{\frac{-R}{\xi_c}}
\label{WSC}
\end{equation}
which decays as a function of $R$ with the length scale $\xi_c$. In the vicinity of the transition the correlation decays slower while a sharp decay can be observed away from the transition point. 
\begin{figure}[t]
	\includegraphics[width=4.2cm,height=2.8cm]{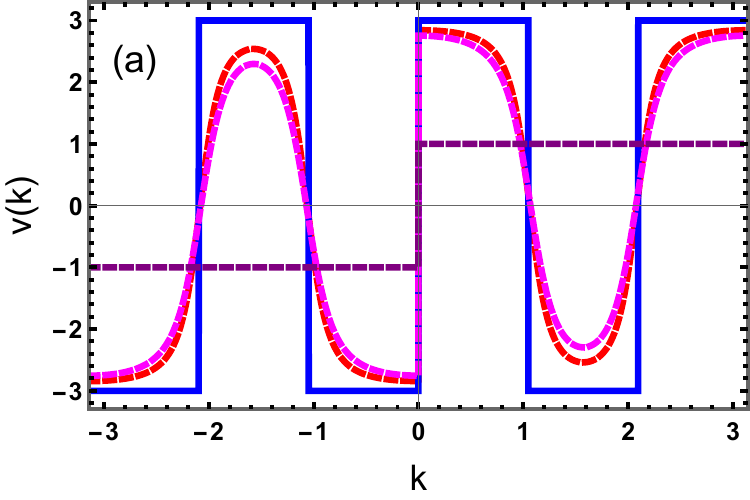}\hspace{0.2cm}\includegraphics[width=4.2cm,height=2.8cm]{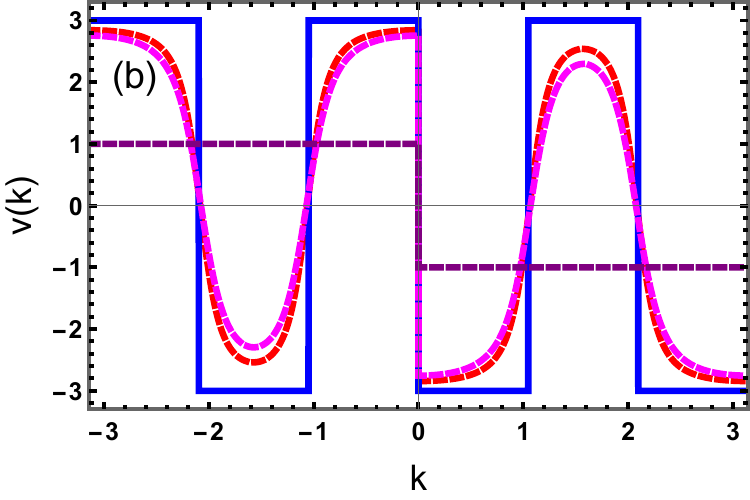}\\
	\includegraphics[width=4.2cm,height=2.8cm]{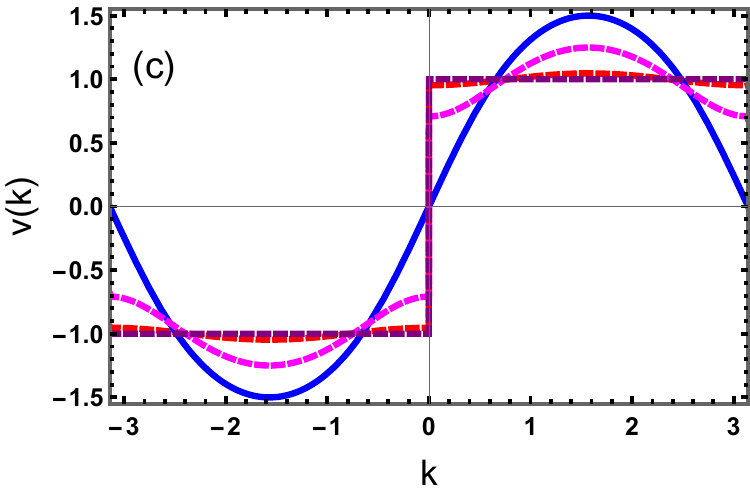}\hspace{0.2cm}\includegraphics[width=4.2cm,height=2.8cm]{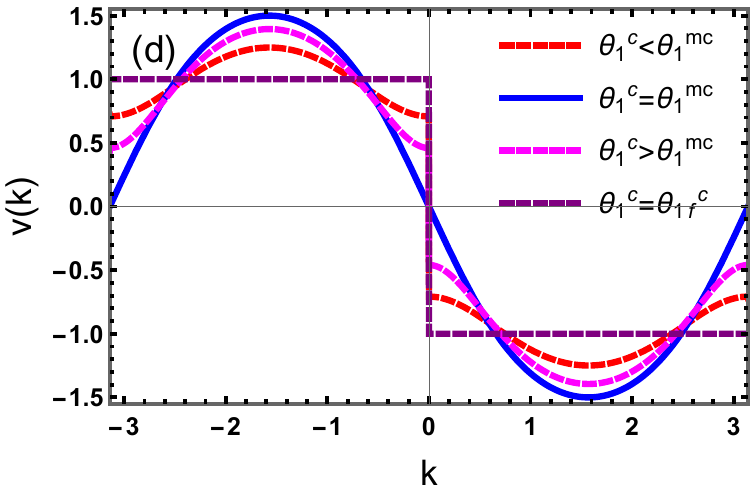}\\
	\caption{Group velocity at and away from multicritical points (The plot legends in (d) is applicable to all the figures). Group velocity for linear dispersive multicritical points on (a) red ($\theta_1^{mc}=0$) and (b) blue ($\theta_1^{mc}=\pm \pi$) critical lines. In both the cases $v(k)$ changes abruptly and span within $[-3,3]$. Group velocity for quadratic dispersive multicritical points on (c) red ($\theta_1^{mc}=\pm \frac{2\pi}{3}$) and (d) blue ($\theta_1^{mc}=\pm \frac{\pi}{3}$) critical lines. Here $v(k)$ changes smoothly and span within $[-1.5,1.5]$. Away from these points $v(k)$ shows abrupt change in sign at the gap closing points and vary non-linearly with $k$. Constant $v(k)$ is retained at the fixed points at $\theta_{1f}^{c}=0,\pm\frac{\pi}{2},\pm \pi$. }
	\label{gv}
\end{figure}

\begin{figure*}[t]
	\includegraphics[width=4.5cm,height=3.5cm]{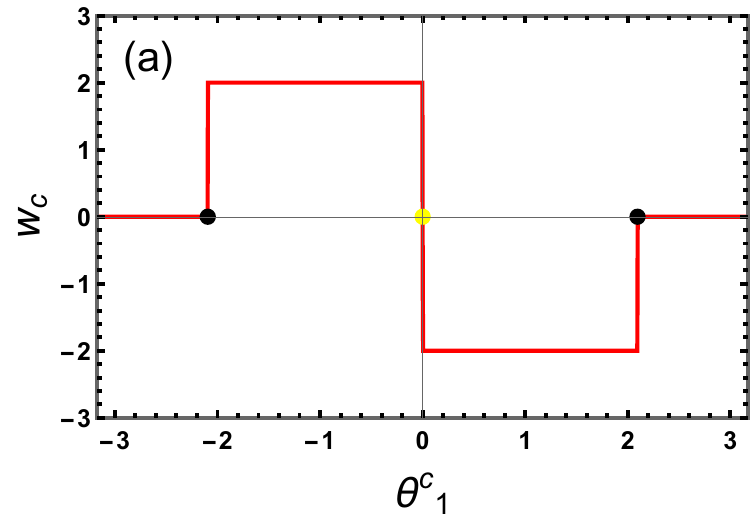}\hspace{0.7cm}\includegraphics[width=4.5cm,height=3.5cm]{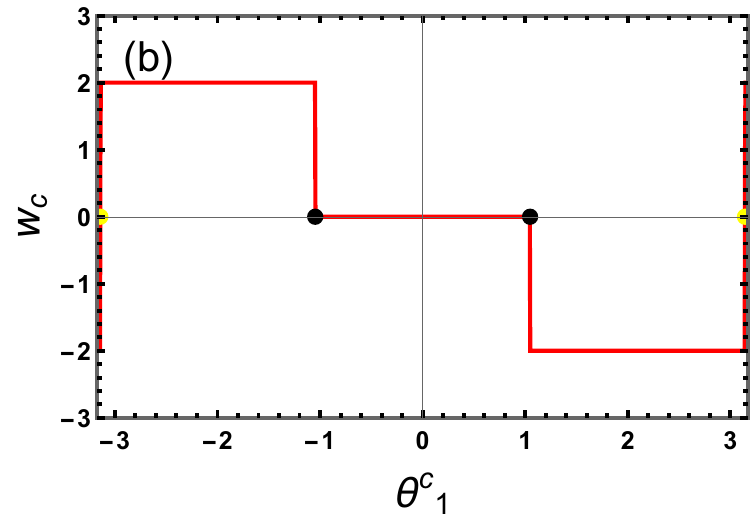}\hspace{0.7cm}\includegraphics[width=4.5cm,height=3.5cm]{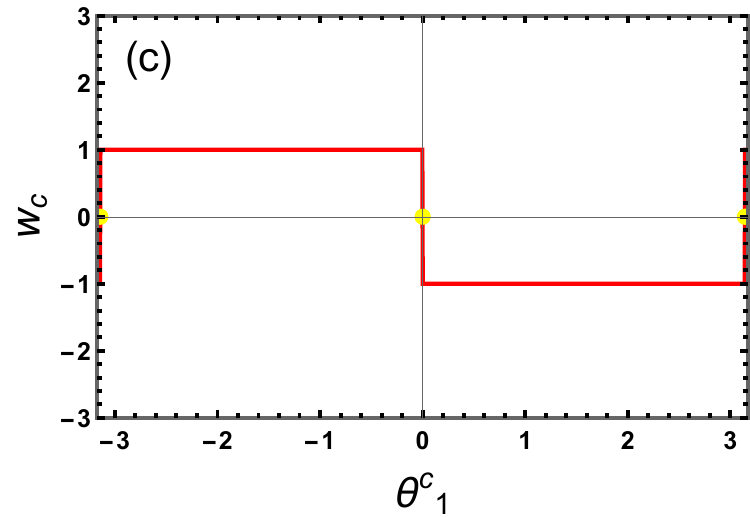}
	\caption{Winding number at various critical lines of three-step quantum walk. Varying $\theta_1^{c}$, the system shows $w_c=0$ and $w_c\neq0$ gapless phases.  (a) The red critical lines in Eq.\ref{R1}-Eq.\ref{R3} shows transitions between $w_c=0$ gapless phases and $w_c=\pm 2$ phases. (b) The blue critical lines in Eq.\ref{B1}-Eq.\ref{B2} shows transitions between $w_c=0$ and $w_c=\pm 2$ gapless phases. (c) Orange and purple critical lines in Eq.\ref{OP1}-Eq.\ref{OP3} remains in $w_c=\pm1$ in the gapless phases.} 
	\label{WC}
\end{figure*}

\begin{figure*}[t]
	\includegraphics[width=5.5cm,height=3cm]{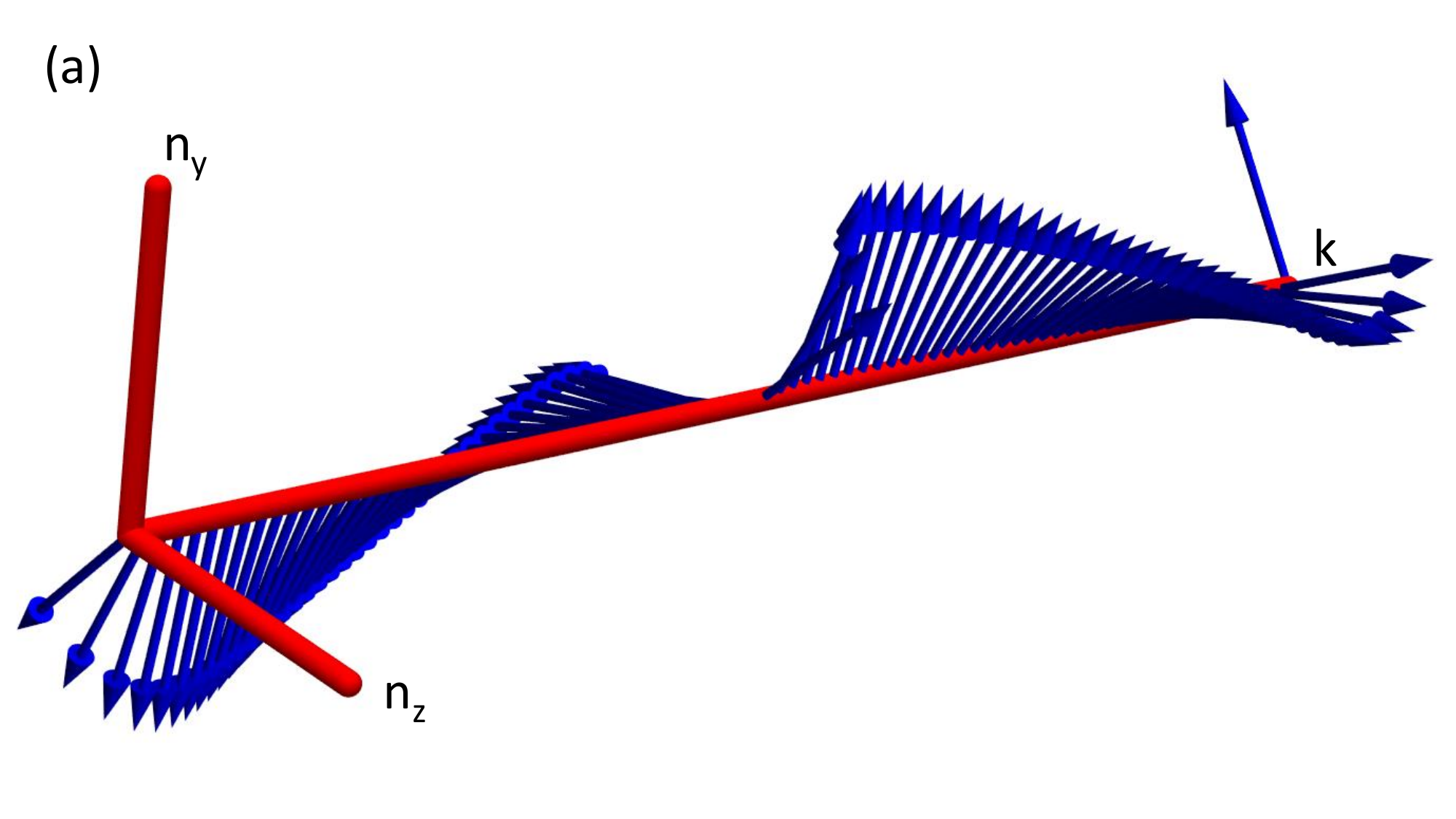}\hspace{0.4cm}\includegraphics[width=5.5cm,height=3cm]{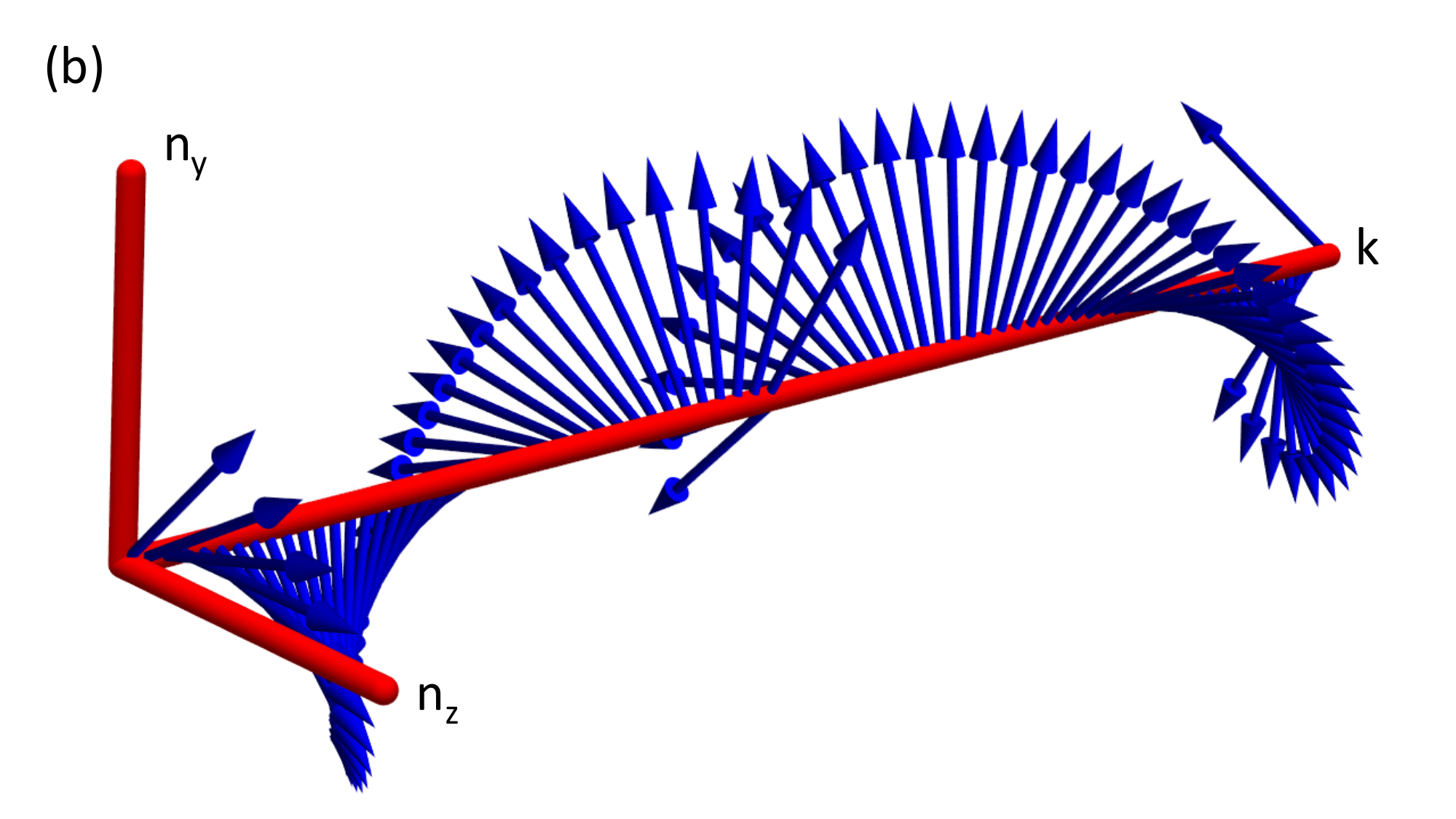}\hspace{0.4cm}\includegraphics[width=5.5cm,height=3cm]{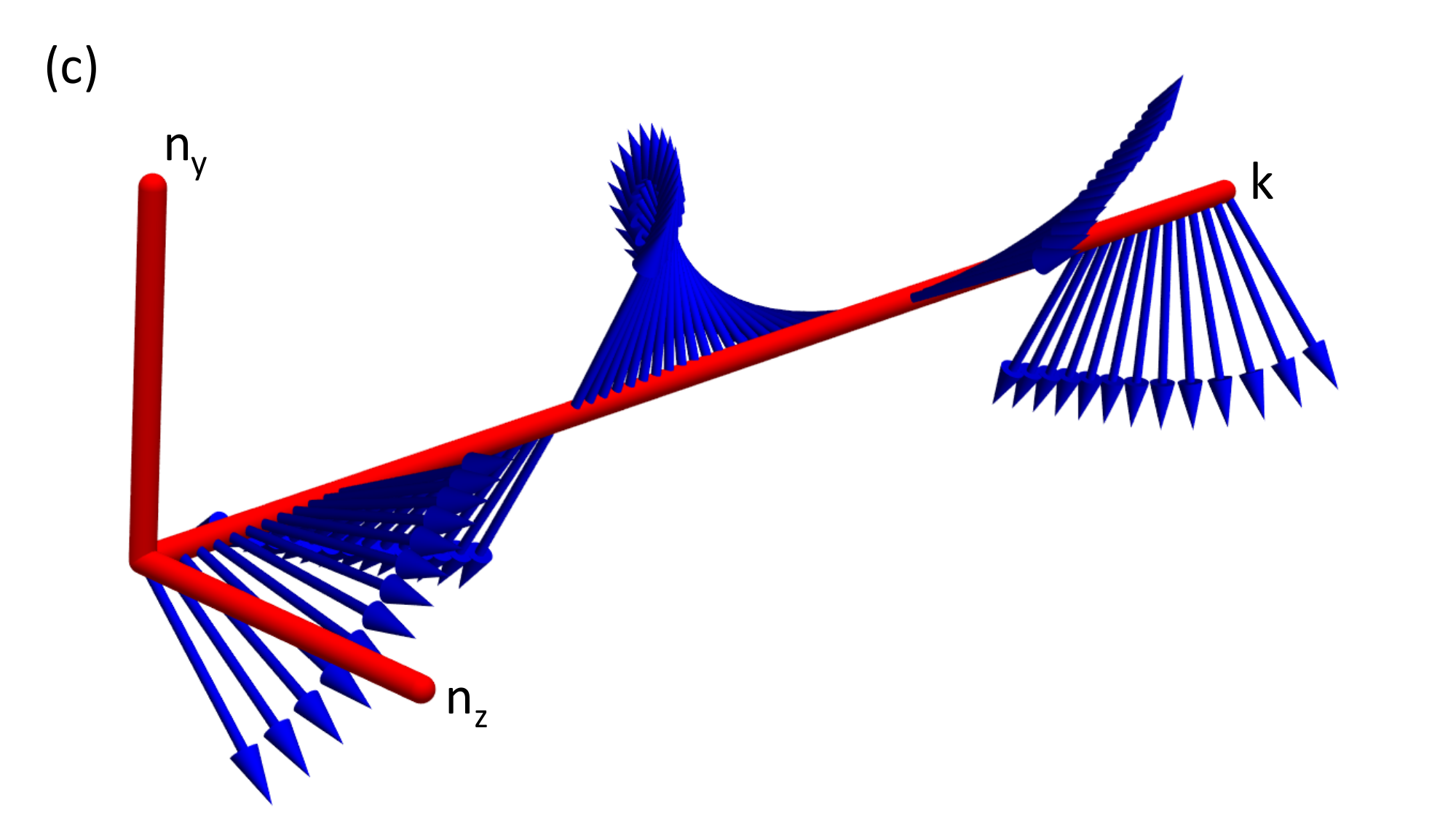}
	\caption{Winding vector $\mathbf{n(k)}$ at various gapless phases of three-step quantum walk.  (a) Shows $w_c=0$ gapless phases and (b) shows $w_c=\pm2$ gapless phases, on both red and blue (high symmetry) critical lines. (c) Shows $w_c=\pm1$ gapless phases for orange and purple (non-high symmetry) critical lines.}
	\label{WCA}
\end{figure*}

We consider all the critical lines and perform RG scheme discussed above to identify the topological transition between gapless phases. The RG equations in Eq.\ref{CRG} for the criticalities in Eq.\ref{R1}-Eq.\ref{R3}, Eq.\ref{B1}-Eq.\ref{B2} and Eq.\ref{OP1}-Eq.\ref{OP3} are obtained respectively as 
\begin{align}
\frac{d\theta_1^{c}}{dl} &= \left( 3\kappa_1-\frac{\kappa_1}{1+2\kappa_2}-\frac{2\kappa_1}{2+\kappa^{r}}\right)+\kappa_3 \label{RG1}\\
\frac{d\theta_1^{c}}{dl} &= \left( \kappa_1+\frac{\kappa_1}{1-2\kappa_2}-\frac{6\kappa_1}{4-\kappa^{b}}\right)+\kappa_3\label{RG2}\\
\frac{d\theta_1^{c}}{dl} &= 2 \left( \frac{\kappa_2}{\kappa_1}\right) \label{RG3}.
\end{align}
where $\kappa_1=\sin(\theta_1^{c})$, $\kappa_2=\cos(\theta_1^{c})$, $\kappa^{r}=\cos(2\theta_1^{c})$, $\kappa^{b}=4\cos(\theta_1^{c})-\cos(2\theta_1^{c})$ and $\kappa_3=\sin(2\theta_1^{c})$. Behavior of these RG equations with $\theta_1^{c}$ is shown Fig.\ref{RG}. As it can be seen evidently from Fig.\ref{RG}(a) that the topological transition at the multicritical points at $\theta_1^{mc}=\pm\frac{2\pi}{3}$ (black dots) can be identified with the diverging flow i.e. $d\theta_1^{c}/dl\rightarrow \infty$ of Eq.\ref{RG1}. Therefore, the curvature function RG  effectively captures the transition between gapless phases on the criticality in Eq.\ref{R1}-Eq.\ref{R3}. Moreover, several fixed points (magenta dots) are obtained at $\theta_{1f}^{c}=0,\pm\frac{\pi}{2},\pm \pi$ where we observe $d\theta_1^{c}/dl\rightarrow 0$. Similarly, the transition between gapless phases on the critical line in Eq.\ref{B1}-Eq.\ref{B2}, i.e. Eq.\ref{RG2} can be observed in Fig.\ref{RG}(b). The diverging flow are identified at the multicritical points $\theta_1^{mc}=\pm\frac{\pi}{3}$ and the fixed points are obtained at $\theta_{1f}^{c}=0,\pm\frac{\pi}{2},\pm \pi$. 

Interestingly, the transition between non-high symmetry gapless phases in Eq.\ref{OP1}-Eq.\ref{OP3} can also be captured from the high-symmetry points RG scheme. In this case, the topological transition occurs only at $\theta_1^{mc}=\pm\pi$, as discussed in the previous section. The diverging flow of Eq.\ref{RG3} is obtained at the multicritical points  at $\theta_1^{mc}=\pm\pi$ while the fixed points are identified at $\theta_{1f}^{c}=\pm\frac{\pi}{2}$, as shown in Fig.\ref{RG}(c). The RG fails to capture the multicritical points at $\theta_1^{mc}=\pm\frac{2\pi}{3},\pm \frac{\pi}{3}$ on these critical lines since they do not host the transition between gapless phase. 

The correlation function of the Wannier states in Eq.\ref{WSC} can be calculated for all the multicritical points which ensures that these points indeed host the transition between various gapless phases. In Fig.\ref{WCF}(a) we shows the correlation function of the Wannier states for quadratic multicritical points at $\theta_1^{mc}=\pm\frac{2\pi}{3},\pm\frac{\pi}{3}$. We observe the decay of $\lambda_c(R)$ is slower close ($\delta\theta_1^{c}=0.1$) to the transition points while it decays faster away ($\delta\theta_1^{c}=0.3$) from these points. In these cases $\lambda_c(R)$ is obtained for the $k_0=0$. However, it can also be obtained for $k_0=\pm \pi$ in which case the $\lambda_c(R)$ shows oscillatory decay similar to the previous case. 

For the linear multicritical points at $\theta_1^{mc}=0,\pm\pi$ we observe a sharp decay close ($\delta\theta_1^{c}=0.1$) to the points and it gets slower away ($\delta\theta_1^{c}=0.3$) from the points, as shown in Fig.\ref{WCF}(b). This due to the fact that these multicritical points also manifests as fixed points as seen from the RG analysis previously. Therefore, the characteristic length $\xi_c\rightarrow0$ as the transition points are approached. This leads to the reduced correlations near to these points. Such behavior of multicritical points is observed in a fermionic system with gapless-gapless transition \cite{kumar2021topological}. Nonetheless, both RG and correlation function of the Wannier states both distinctly characterize the topological transition between gapless phases at the multicritical points at various criticalities.

\subsection{Group velocity}
Conventional topological transition between gapped phases in quantum walks can be characterized using group velocity of the energy eigenstates \cite{panahiyan2019,panahiyan2020}
\begin{equation}
v(k)=\frac{dE}{dk}
\end{equation}
We show that the transitions at the multicritical points between gapless phases are also recognized using group velocity. However, in this case the behavior of the $v(k)$ is different from the conventional case. The group velocity of the three-step quantum walk considered in this work can be obtained using the energy dispersion in Eq.\ref{eigen} as
\begin{equation}
v(k)= \pm\frac{\alpha^{\prime}}{\sqrt{1-\alpha}}
\end{equation}
where 
\begin{align}
\alpha&=\cos(3 k) \cos(\theta_1) \cos(\theta_2)^2 \nonumber \\&- 
\cos(k) \sin(\theta_2) (2 \cos(\theta_2) \sin(\theta_1) + 
\cos(\theta_1) \sin(\theta_2)), \nonumber	
\end{align}
and $\alpha^{\prime}$ is the first derivative.

Group velocity at two kinds of multicritical points (linear and quadratic) is shown in Fig.\ref{gv}. For linear dispersive multicritical points, the $v(k)$ changes abruptly from positive to negative or vice versa at the gap closing $k_0$ points in the Brillouin zone, as shown in Fig.\ref{gv}(a) and (b). Moreover, it remains constant and spans within $[-3,3]$ on either side of the gap closing points. In contrast, for quadratic multicritical points $v(k)$ changes smoothly and $v(k)=0$ at $k_0$ points, as shown in Fig.\ref{gv}(c) and (d). It remains non-linear with $k$ away from the gap closing points and spans within $[-1.5,1.5]$. This is due to the quadratic nature of the dispersion at the gap closing points, for which the slope changes smoothly and flips sign. 

The multicritical points at different (red and blue) critical lines can be identified from the sign of $v(k)$ as $k$ traverse through $[-\pi,0)$ and $(0,\pi]$. For quadratic phase transition points on red and blue critical lines, the sign of $v(k)$ changes 
for $k$ through $[-\pi,0)$ and $(0,\pi]$. It turns zero at the gap closing high-symmetry points, as shown in Fig.\ref{gv}(c) and (d). For linear phase transition points, the gap closes at both high and non-high symmetry points with $v(k)$ abruptly changing sign at each of these points. However, the signs of $v(k)$ for all $k$ through $[-\pi,0)$ is therefore opposite to the corresponding $k$ through $(0,\pi]$ between red and blue critical lines, as shown in Fig.\ref{gv}(a) and (b). Moreover, as shown in Fig.\ref{gv}, away from the multicritical points along the critical lines, group velocity shows an abrupt change at gap closing points; however, it does not remain constant on either side. Therefore, at the distinct gapless phases $v(k)$ changes non-linearly with $k$ except at some special point where linear variation is retained.  These points turn out to be the fixed points of the gapless phases at $\theta_{1f}^{c}=0,\pm\frac{\pi}{2},\pm \pi$ as found from the previous section of RG analysis (see Fig.\ref{RG}). Such behavior of $v(k)$ is due to the fact that the dispersion varies completely linearly even away from the Dirac points at both $E=0,\pm\pi$. Away from the fixed and linear multicritical points, the dispersion varies linearly only near the Dirac points. 

\section{Topological character of gapless phases}\label{sec5}
In this section we identify the topological characters associated with the various gapless phases on the critical lines in Eq.\ref{R1}-Eq.\ref{OP3}. The topological characters of gapped phases can be captured from the topological invariant numbers defined in Eq.\ref{winding}. However, the winding number becomes ill-defined at the gapless points \cite{verresen2018topology}. Nonetheless, the topological characters of the gapless phases can be captured using the near critical approach as discussed in the previous section. This approach avoids the singularities of an exact gapless point and enables one to define the winding number associated with the various gapless phases of the quantum walk. This winding number encapsulate the information of both criticality and topology \cite{kumar2021topological} of the quantum walk.
 
Now one can redefine the winding number in Eq.\ref{winding} for gapless phases using the critical line relations in Eq.\ref{R1}-Eq.\ref{OP3}. Therefore, one can write
\begin{equation}
w_c=\frac{1}{2\pi}  \lim_{\delta \rightarrow 0}\oint\limits_{|k-k_0|>\delta} \frac{d\theta^c_k}{dk} dk,
\end{equation}
where $\theta^c_k=\tan^{-1}(d^c_3/d^c_2)$. Depending on the components $d^c_3$ and $d^c_2$ for the critical lines in Eq.\ref{R1}-Eq.\ref{OP3}, the corresponding winding number $w_c$ shows the topological character of the gapless phases on that critical lines. 


At first, we consider the red critical lines in Eq.\ref{R2}-Eq.\ref{R3} and calculate $w_c$ for varying $\theta_1^{c}$, as shown in Fig.\ref{WC}(a). As $\theta_1^{c}$ is varied from $-\pi$, we identify a gapless phase with $w_c=0$ which undergoes a transition into a gapless phase with $w_c=2$ then to $w_c=-2$ and again back to $w_c=0$. These transitions occurs at a multicritical point $\theta_1^{c}=-\frac{2\pi}{3}, 0$ and $\frac{2\pi}{3}$ respectively. 
For the blue critical lines in Eq.\ref{B1}-Eq.\ref{B2}, a similar transition between gapless phases with $w_c=0$ and $w_c=\pm 2$ can be observed as shown in Fig.\ref{WC}(b). These transitions occur at the mulricritical points $\theta_1^{c}=\pm\frac{\pi}{3}, \pm\pi$.  
Moreover, the gapless phases on the non-high symmetry critical lines i.e. orange and purple lines in Eq.\ref{OP1}-Eq.\ref{OP3}, can be identified with $w_c=\mp1$, as shown in Fig.\ref{WC}(c). However, these values are valid only for the range of $\theta_1^{c}$ in which bulk gap is closed. The transition occurs only at the points $\theta_1^{c}=0,\pm\pi$ and other multicritical points $\theta_1^{c}=\pm\frac{2\pi}{3}, \pm\frac{\pi}{3}$ on these lines are not captured by the winding number. This also indicate that there is no gapless-gapless transition at these points. 

We also look at the winding of unit vector $\mathbf{n(k)}$ in the Brillouin zone and verify the topological characters associated with the various gapless phases. For gapless phases with $w_c=0$, the winding vector $\mathbf{n(k)}$ shows incomplete windings along the $k$ axis, 
as shown in Fig.\ref{WCA}(a). However, Fig.\ref{WCA}(b) shows 
gapless phase with $w_c=\pm2$ and Fig.\ref{WCA}(c) shows $w_c=\pm1$ gapless phases where $\mathbf{n(k)}$ shows two and one winding with varying $k$, respectively. 

\section{Summary and Conclusion}\label{sec6}
In this work, we have shown that the topological gapless phases and the unique transition between them can be simulated in a discrete-time three-step quantum walk. 
Implementing a topological chain in a three-step discrete-time quantum walk, we realize 
various gapless phases with distinct topological properties. These gapless phases can be identified with pairs of high symmetry and non-high symmetry gap closing points. The intersection of both kinds of criticalities are 
identified as multicritical points with either quadratic or linear dispersions. Both kinds of multicritical points mark the unique topological transition between distinct gapless phases without gap opening and closing.  

The topological transitions have been characterized by reconstructing the scaling theory based on the curvature function using near near-critical approach for the quantum walk. Curvature function shows diverging peak and flips the sign at the multicritical points indicating the topological transitions. The peak develops either at high symmetry points or non-high symmetry points and shows swaping properties between them. The critical exponents of the diverging curvature function obey usual sacling law $\gamma=\nu$ for one dimensional systems. Morever, the renormalization group flow based on the curvature function diverge at the multicritical points and vanish at fixed points. The Wannier state correlation function shows the slower decay near the multicritical points indicating the topological transition between gapless phases. Besides, we also realize the instances of overlapping of critical and fixed point properties, as in the topological chains, at linear dispersive multicritical points. The phase transitions at various multicritical points are further characterized using the group velocity of energy eigenstates. 
Moreover, the topological characters of various gapless phases and the transitions are identified from the winding number defined at the criticalities using near near-critical approach.  
We have observed that a gapless phase possesses topological characters identified with different winding numbers. We have identified $w_c=\pm2$, $w_c=\pm1$ and $w_c=0$ gapless phases and the transition between them at multicritical points. The winding vector at these gapless phases also shows two, one and zero complete windings accordingly. 

Therefore, topological gapless phases and the all intriguing features of the unique transition transition between them in topological chain can be simulated in quantum walks. However, the quantum walk considered in this work is unitary and therefore simulates the Hermitian topological chains. The non-unitary version of the three-step quantum walk can simulate non-hermitian counterparts of the topological chains \cite{hideakimodel,tian}. Therefore, an extension of this work in non-unitary quantum walks can allow one to study the fate of these gapless phases in non-hermitian systems. This sets the future direction of this work.

\begin{acknowledgements}
	RRK acknowledges the Indian Institute of Technology (IIT), Bombay for support through the Institution Post-Doc program.
\end{acknowledgements}

\end{document}